\documentclass[reprint,twocolumn]{revtex4}
\usepackage{amsfonts}
\usepackage{amsmath}
\usepackage{amssymb}
\usepackage{charter}
\usepackage{graphicx}
\usepackage{float}

\begin{document}

\title{Improving phase estimation using the number-conserving operations}
\author{Huan Zhang,$^{1}$}
\author{Wei Ye,$^{1,2}$}
\thanks{yeweicsu@csu.edu.cn}
\author{Chaoping Wei,$^{3}$ Cunjin Liu,$^{1}$}
\author{Zeyang Liao $^{4}$}
\thanks{liaozy7@mail.sysu.edu.cn}
\author{Liyun Hu$^{1}$}
\thanks{hlyun@jxnu.edu.cn}
\affiliation{$^{{\small 1}}$\textit{Center for Quantum Science and Technology, Jiangxi
Normal University, Nanchang 330022, China}\\
$^{{\small 2}}$\textit{School of Computer Science and Engineering, Central
South University, Changsha 410083, China}\\
$^{{\small 3}}$\textit{Key Laboratory of Water Information Cooperative
Sensing and Intelligent Processing, Nanchang Institute of Technology,
Nanchang 330022, China}\\
$^{{\small 4}}$\textit{School of Physics, Sun Yat-sen University, Guangzhou
510275, China}}

\begin{abstract}
We propose a theoretical scheme to improve the resolution and precision of
phase measurement with parity detection in the Mach-Zehnder interferometer by using a nonclassical input state which is
generated by applying a number-conserving generalized superposition of
products (GSP) operation, $\left( saa^{\dagger }+ta^{\dagger }a\right) ^{m}$
with $s^{2}+t^{2}=1$, on two-mode squeezed vacuum (TMSV) state. The nonclassical properties of the
proposed GSP-TMSV are investigated via average photon number (APN),
anti-bunching effect, and degrees of two-mode squeezing. Particularly, our results show that both
higher-order $m$ GSP operation and smaller parameter $s$ can increase the total APN, which leads to the improvement
of quantum Fisher information. In addition, we also compare the phase
measurement precision with and without photon losses between our scheme and
the previous photon subtraction/addition schemes. It is found that
our scheme, especially for the case of $s=0$,  has the best performance via the enhanced phase resolution and
sensitivity when comparing to those previous schemes even in the presence of photon
losses. Interestingly, without
losses, the standard quantum-noise limit (SQL) can always be surpassed in our
our scheme and the Heisenberg limit (HL) can be even achieved when $s=0.5,1$ with small total APNs.
However, in the presence of photon losses, the HL cannot be beaten, but the
SQL can still be overcome particularly in the large total APN regimes. Our results  here can find important applications in quantum metrology.

\textbf{PACS: }03.67.-a, 05.30.-d, 42.50,Dv, 03.65.Wj
\end{abstract}

\maketitle

\section{Introduction}

The ultimate aim of quantum metrology is to achieve a higher precision and
sensitivity of the phase estimation using (non)classical field of light as
the input of optical interferometers \cite{1,2,3,4}. Among them, the
Mach-Zehnder interferometer (MZI) is one of the most practical
interferometers, and its phase sensitivity is limited by the standard
quantum-noise limit (SQL) $\Delta \varphi =1/\sqrt{N}$ ($N$ is the average
number of photons inside the interferometer), together with solely classical
resources as the input of the MZI \cite{5}. In order to go beyond this
limit, both the nonclassical states \cite{6,7} and the entangled states \cite%
{2,8,9} are applied to quantum metrology, which results in the reduction of
the phase uncertainty, thereby reaching the Heisenberg limit (HL) $\Delta
\varphi =1/N$ \cite{10}. For instance, Dowling \textit{et al.} \cite{2}
pointed out that the so-called N00N states in quantum optical interferometry
can achieve the HL. Unfortunately, these states are extremely sensitive to
photon losses \cite{9,10,11}. To solve this problem, Anisimov \textit{et al.}
\cite{8} theoretically studied that using the two-mode squeezed vacuum state
(TMSV) as the input of the MZI with parity detection scheme can reach the
so-called sub-Heisenberg limit with small total average photon numbers
(APN). However, restricted by current experimental techniques, it is still
difficult to generate strongly entangled TMSV in which its maximum
obtainable degree is about $r=1.15$ $(\overline{n}=\sinh ^{2}r\approx 2)$
\cite{14}. Thus, how to prepare highly non-classical and strongly entangled
quantum states has become one of the most important topics in quantum information and quantum metrology.

For this purpose, the usage of non-Gaussian operations \cite%
{15,16,17,18,19,20,21,22,23,24} is a feasible method, e.g., photon
subtraction (PS) \cite{15}, photon addition (PA) \cite{18,19,20,21,22}, and
their superposition \cite{23,24}, which also plays an vital role in quantum
illumination \cite{25,26}, quantum cryptography \cite{27,28,29,30,31} and
quantum teleportation \cite{32,33,34}. For instance, Agarwal and Tara
proposed that the classical coherent states can be transformed into highly nonclassical quantum states by the PA operation
\cite{18}
and this PA operation can be experimentally  implemented which was proposed by Zavatta
\cite{19}. In addition, highly nonclassicality has been shown for the PA- (or PS-) squeezed states \cite{35,36}. Based on the facts
mentioned above, Gerry \textit{et al.} \cite{6} first proposed to use the PS-TMSV (simultaneously subtracting the same number of photons from the
TMSV ) as the input of the MZI, and showed that the phase
measurement uncertainty of the PS-TMSV scheme is smaller than that of the usual
TMSV for the same squeezing parameters. Then, Ouyang \textit{et al.}
\cite{22} used the PA-TMSV as the input state of the MZI, and  showed that it has better performance in terms of phase sensitivity for small phase shift when compared with
both the PS-TMSV and the usual TMSV. In addition to the
aforementioned typical non-Gaussian operations, here we suggest to use a new type of non-Gaussian states as the input of the MZI in an attempt
to further enhance the resolution and sensitivity of the phase estimation. The non-Gaussian states we consider here are the output states by applying  the number-conserving generalized superposition of
products (GSP) operation $\left( saa^{\dagger }+ta^{\dagger }a\right)$ with $%
s^{2}+t^{2}=1$, to the TMSV.
It is interesting to notice that the PA-then-PS ($aa^{\dagger }$) and the PS-then-PA ($a^{\dagger }a$) as well as their superposition can be considered as three special cases in our scheme. In particular, the first two have been used to improve entanglement and fidelity of quantum teleportation, but none of them are used to improve phase measurement accuracy.
Not only can this GSP operation be implemented experimentally, proposed by
Kim \cite{37}, but also the GSP operation on the TMSV is able to generate a
strongly entangled non-Gaussian state as well \cite{38,39}.

In order to extract quantum phase information more effectively,  three types of detection schemes are usually used, including
intensity detection \cite{40,41}, homodyne detection \cite{42} and parity
detection \cite{43,44}. It should be noted that not all detection schemes
can employ the full potential of nonclassical states to achieve the
superresolution and supersensitivity. In particular, as referred to Ref.
\cite{45}, the intensity detection is more suitable for optical
interferometers with coherent light as input, but it is not applicable to
the TMSV. In contrast, the parity detection can be used in the quantum metrology with the TMSV to achieve the superresolution and even sub-Heisenberg limit sensitivity \cite{8, 46,47}.
Thus, in this
paper, we take advantage of parity detection to extract phase information and study the phase resolution and sensitivity of the MZI by
using the GSP-TMSV as input. The numerical simulation results show that
our scheme, especially for the case of the PS-then-PA TMSV ($s=0$), is
always superior to the original TMSV scheme in terms of the quantum
Fisher information (QFI) and the phase resolution and sensitivity.
Dramatically, the SQL can be  always surpassed in our scheme and the HL can even be beaten  for
the cases when $s=0.5,1$ in the regime of
small total APN. Furthermore, since the interaction with the environment is inevitable,  we also investigate the effects of GSP operations against the photon
losses placed in front of parity detection (denoted as an external loss) and
between the phase shifter and the second beam splitter (BS) (denoted as an
internal loss) from a practical point of view.
Our results show that in the presence of photon losses the phase sensitivity
with the GSP-TMSV, especially for the case of $s=0$, can still be  better
than that with both the TMSV and the PA(PS)-TMSV under the same accessible
parameters. Interestingly, we also find that the effects of the external
losses on phase uncertainty are more serious than the internal-loss cases.

The structure of this paper is organized as follows: In Sec. II, we briefly outline
the preparation of the GSP-TMSV state, and then present its nonclassicality
according to APN, antibunching effect and two-mode squeezing property. In
Sec. III, we show the application of the GSP-TMSV in the MZI and mainly
focus on its QFI behavior. The resolution and sensitivity of phase
estimation with parity detection are further discussed in Sec. IV.  In Sec.
V, we mainly pay attention to the effects of photon losses, involving
external and internal losses, on the resolution and sensitivity. Finally,
the main results are summarized in Sec. VI.

\section{The generation of the GSP-TMSV and nonclassical properties}

In this section, we first introduce the GSP-TMSV in theory, and then show
its nonclassicality by means of APN, anti-bunching effect and two-mode
squeezing property.

\subsection{The generation of the GSP-TMSV}

\begin{figure}[tbp]
\label{Fig1} \centering \includegraphics[width=8cm]{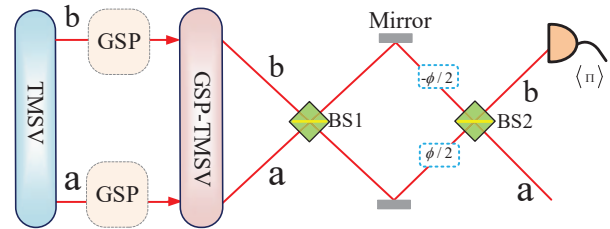}
\caption{{}(Color online) Schematic diagram of a balanced MZI for the
detection of the phase shift (Violet) when the GSP-TMSV state is sent to the
first BS (Green), and the photon-number parity measurements are performed on
the output $b$ mode.}
\end{figure}
\

In recent years, it has been demonstrated that both the PS-TMSV and the
PA-TMSV as the inputs of the MZI can improve the phase sensitivity
effectively \cite{6,22}, since these nonGaussian states have the advantages
over the Gaussian states in terms of the nonclassicality and the
entanglement degree. In this section, we introduce a new kind of
non-Gaussian state, the GSP-TMSV, which can be prepared by acting two GSP
operations on the TMSV, as pictured in Fig.1 (orange box). As referred to
\cite{38,39}, this GSP operation can be seen as an equivalent operator
\begin{equation}
\hat{O}=\left( s_{1}aa^{\dagger }+t_{1}a^{\dagger }a\right) ^{m}\left(
s_{2}bb^{\dagger }+t_{2}b^{\dagger }b\right) ^{n},  \label{1}
\end{equation}%
where $s_{i}^{2}+t_{i}^{2}=1\ (i=1,2)$ and both $a$ $\left( a^{\dagger
}\right) $ and $b$ $\left( b^{\dagger }\right) $ are annihilation (creation)
operators for modes $a$ and $b$, respectively. Note that $\left( m,n\right) $
represent $m$-order operation of $s_{1}aa^{\dagger }+t_{1}a^{\dagger }a$ on
mode $a$ and $n$-order operation of $s_{2}bb^{\dagger }+t_{2}b^{\dagger }b$
on mode $b$. Thus, the GSP-TMSV can be given by
\begin{eqnarray}
\left \vert \psi \right \rangle _{ab} &=&\frac{\hat{O}S_{2}\left( z\right) }{%
\sqrt{P_{d}}}\left \vert 00\right \rangle  \notag \\
&=&\frac{\Re u}{\sqrt{P_{d}}}\exp \left( va^{\dagger }b^{\dagger }\right)
\left \vert 00\right \rangle ,  \label{2}
\end{eqnarray}%
with%
\begin{eqnarray}
\Re &=&\frac{\partial ^{m+n}}{\partial \tau _{1}^{m}\partial \tau _{2}^{n}}%
\left \{ \cdot \right \} |_{\tau _{1}=\tau _{2}=0},  \notag \\
u &=&\sqrt{1-z^{2}}\exp \left( s_{1}\tau _{1}+s_{2}\tau _{2}\right) ,  \notag
\\
v &=&z\exp \left( s_{1}\tau _{1}+t_{1}\tau _{1}+s_{2}\tau _{2}+t_{2}\tau
_{2}\right) ,  \label{3}
\end{eqnarray}%
where $S_{2}\left( z\right) =\exp [\left( a^{\dagger }b^{\dagger }-ab\right)
$arctanh $z]$ is the two-mode squeezing operator with a squeezing parameter $%
z$ and $P_{d}$ is a normalization coefficient which can be calculated as%
\begin{equation}
P_{d}=\widetilde{\Re }\frac{uu_{1}}{1-vv_{1}},  \label{4}
\end{equation}%
with
\begin{eqnarray}
\widetilde{\Re } &=&\frac{\partial ^{2m+2n}}{\partial \tau _{1}^{m}\partial
\tau _{2}^{n}\partial \tau _{3}^{m}\partial \tau _{4}^{n}}\left \{ \cdot
\right \} _{|\tau _{1}=\tau _{2}=\tau _{3}=\tau _{4}=0},  \notag \\
u_{1} &=&\sqrt{1-z^{2}}\exp \left( s_{1}\tau _{3}+s_{2}\tau _{4}\right) ,
\notag \\
v_{1} &=&z\exp \left( s_{1}\tau _{3}+t_{1}\tau _{3}+s_{2}\tau _{4}+t_{2}\tau
_{4}\right) .  \label{5}
\end{eqnarray}%
It should be emphasized that for simplicity, all the following simulations
are based on the assumption of $s_{1}=s_{2}=s,t_{1}=t_{2}=t$. In particular,
when $s=0,0.5$ and $1$, from Eqs. (\ref{1}) and (\ref{2}), one can obtain
the PS-then-PA TMSV, a general GSP-TMSV and the PA-then-PS TMSV,
respectively.

For the sake of analysis in the following, here we present the expectation
value of a general quantum operator, i.e.,%
\begin{equation}
\left \langle a^{l}b^{k}a^{\dagger h}b^{\dagger g}\right \rangle =\widetilde{%
\Re }\widetilde{D}P_{d}^{-1}uu_{1}\Delta e^{\Delta w},  \label{6}
\end{equation}%
with%
\begin{eqnarray}
\widetilde{D} &=&\frac{\partial ^{l+k+h+g}}{\partial \tau _{5}^{l}\partial
\tau _{6}^{k}\partial \tau _{7}^{h}\partial \tau _{8}^{g}}\left \{ \cdot
\right \} |_{\tau _{5}=\tau _{6}=\tau _{7}=\tau _{8}=0},  \notag \\
\Delta &=&\left( 1-vv_{1}\right) ^{-1},  \notag \\
w &=&\tau _{7}\tau _{8}v_{1}+\tau _{6}\tau _{5}v+\tau _{6}\tau _{8}+\tau
_{5}\tau _{7},  \label{7}
\end{eqnarray}%
where $l,k,h$ and $g$ are integers ($\geqslant 0$), Eq. (\ref{6}) can be
used to calculate some expectation values, such as $\left \langle
aa^{\dagger }\right \rangle ,$ $\left \langle bb^{\dagger }\right \rangle ,$
$\left \langle aa^{\dagger }bb^{\dagger }\right \rangle ,$ $\left \langle
a^{2}b^{\dagger 2}\right \rangle ,$ and $\left \langle a^{\dagger
2}b^{2}\right \rangle $.

\subsection{Nonclassical\ properties of the GSP-TMSV}

As described in Refs. \cite{6,7}, the nonclassical states of optical field
offer a significant improvement in the sensitivity and precision of the MZI,
thereby promoting the development of quantum metrology. Before investigating
how does the GSP-TMSV as the input affect the sensitivity and resolution of
the MZI, let us first examine its nonclassicality in terms of APN,
anti-bunching effect and two-mode squeezing property, which provide the
basis for the performance improvement of the phase estimation in next
section.

\subsubsection{Average photon number}

As one of statistical\ properties of the light field, the APN is an
important factor for optical interferometry. In addition, as a kind of
non-Gaussian operation, the PS from squeezed vacuum state can surprisingly increase the APN, by which the phase sensitivity can
be improved. Here, we first pay attention to the APN and examine if it can be increased by the GSP
operation or not. According to Eq. (\ref{6}), the APN, say for
mode $a$, can be calculated as
\begin{align}
\overline{N}_{a}& =\left \langle a^{\dagger }a\right \rangle =\left \langle
aa^{\dagger }\right \rangle -1  \notag \\
& =\frac{\widetilde{\Re }uu_{1}}{P_{d}}\frac{\partial ^{2}}{\partial \tau
_{5}\partial \tau _{7}}\Delta e^{\Delta \tau _{5}\tau _{7}}|_{\tau _{5}=\tau
_{7}=0}-1.  \label{8}
\end{align}%
For mode $b$, there is the same result, i.e\textit{.}, $\overline{N}_{a}=%
\overline{N}_{b}=N,$ which can be easily seen from Eq. (\ref{6}).

\begin{figure}[tbp]
\label{Fig2} \centering \includegraphics[width=0.8\columnwidth]{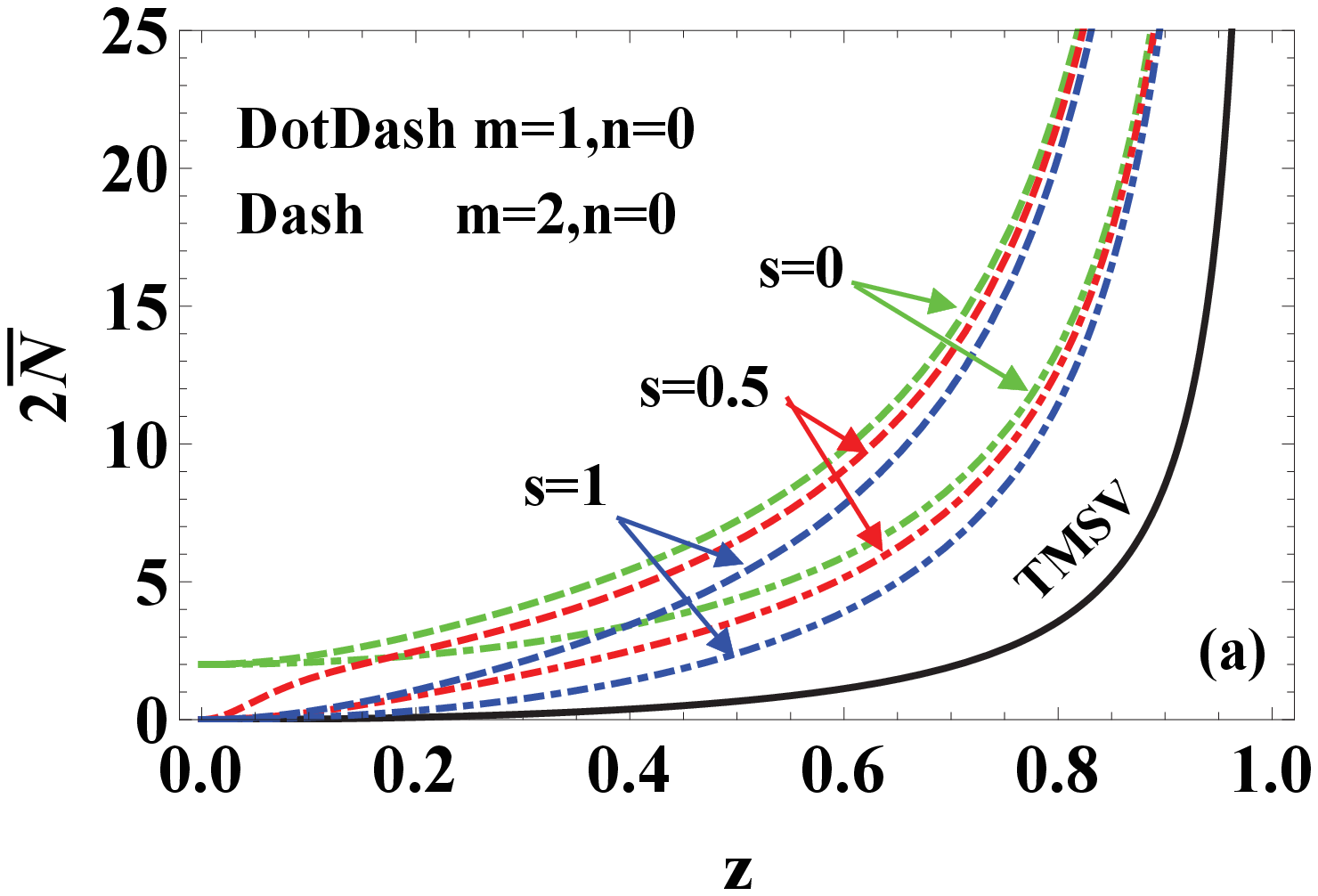} %
\includegraphics[width=0.8\columnwidth]{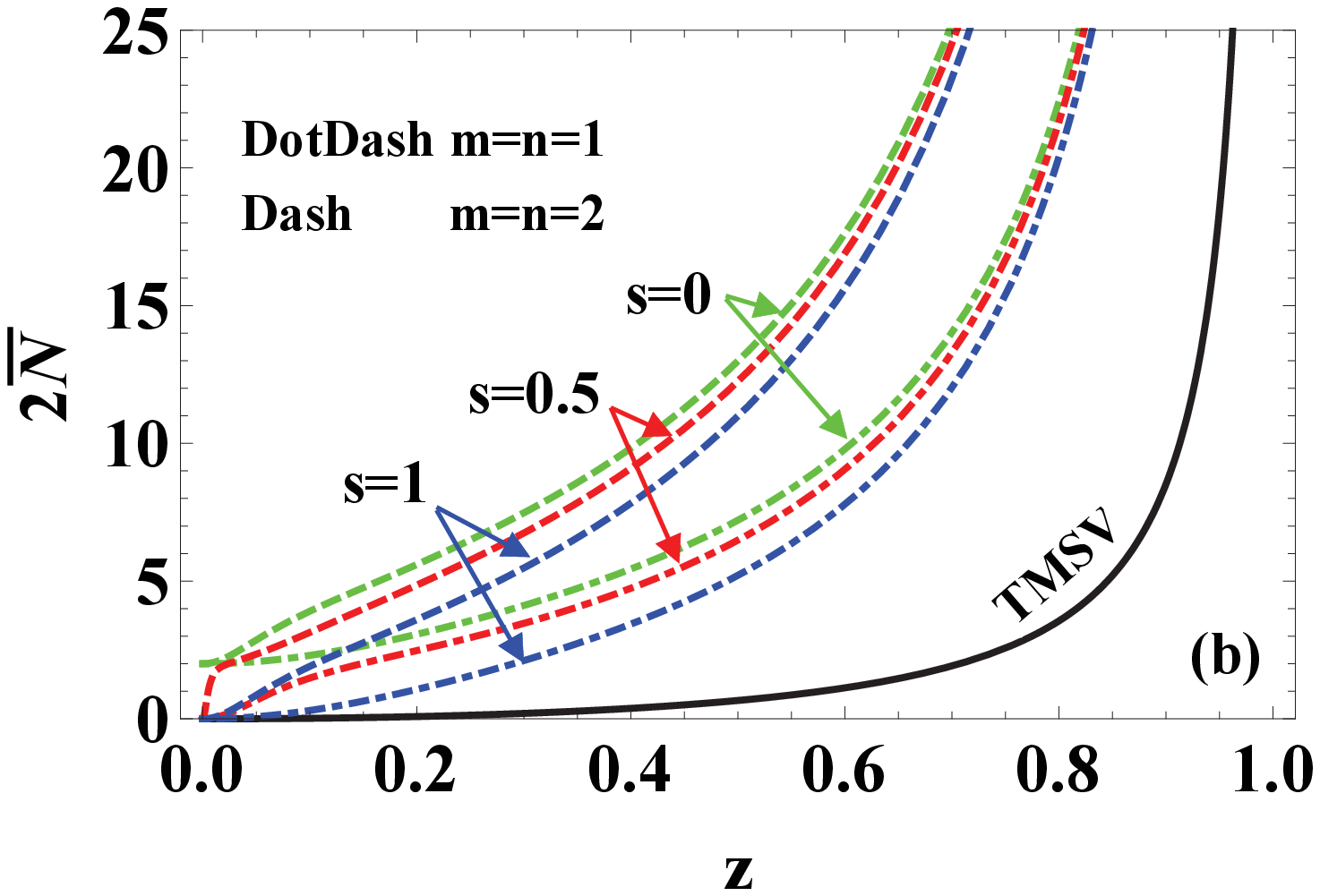}
\caption{{}(Color online) Average photon number as a function of squeezing
parameter $z$ for different operator parameter $s=0,0.5,1$. for (a)
single-side GSP operations ($\left( m,n\right) \in \left \{ \left(
0,1\right) ,\left( 0,2\right) \right \} $), (b) two-side symmetric GSP
operations ($\left( m,n\right) \in \left \{ \left( 1,1\right) ,\left(
2,2\right) \right \} $). Solid lines correspond to the TMSV case.}
\end{figure}

Figure 2 shows the total APN $\left( 2N\right) $ before injecting into the MZI
as the function of the squeezing parameter $z$ for different superposition
parameters $s=0,0.5,1$. For a comparison, the APN of the TMSV is also
plotted in Fig. 2, see the solid black line. From Fig. 2, it is clear that
the APN of the generated states outperforms that of the TMSV in nearly all
squeezing ranges for both single-side and two-side GSP operations. In
addition, for a fixed superposition $s$, the APN increases as the increasing
($m,n$) and $z$. The APN with two-side symmetrical GSP ($\left( m,n\right)
\in \left \{ \left( 1,1\right) ,\left( 2,2\right) \right \} $) is bigger
than that with single-side case ($\left( m,n\right) \in \left \{ \left(
0,1\right) ,\left( 0,2\right) \right \} $) by comparing Fig. 2(a) with 2(b).
On the other hand, it is interesting to notice that, for fixed $m$ and $n$,
the APN decreases as the increasing $s$. In particular, in the limit $s=0,$
corresponding to the PS-then-PA case, the APN has the biggest value when
other parameters are fixed. While for the case of $s=1$ corresponding to the
PA-then-PS case, the APN has the lowest value when comparing with other
cases for $s$. Even so, both PA-then-PS and PS-then-PA have bigger APN than
the TMSV. Among these non-Gaussian operations, the PS-then-PA case presents
the biggest APN.
\begin{figure}[tbp]
\label{Fig3} \centering \includegraphics[width=7.2cm]{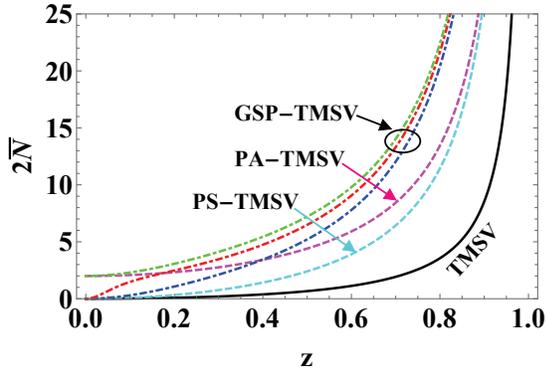}
\caption{{}(Color online) As a comparison, the APN as a function of the
squeezing parameter $z$. The dot-dashed lines represent our scheme for
operation parameters $s=0,0.5,1$ (corresponding to green, red, and blue
color line, respectively), and dashed lines represent the previous work of
using the PA-TMSV (magenta color line) and the PS-TMSV (cyan color line) as
inputs. Solid line corresponds to the TMSV case. }
\end{figure}

In Fig. 3, under the same parameter of $m=n=1$, we also compare about the
APN $2N$ changing with $z$ for giving several non-Gaussian states, including
the PA-TMSV (Magenta dashed), the PS-TMSV (cayan dashed) and the GSP-TMSV.
Distinctly, the APN of the GSP-TMSV is always greater than that of the
PS-TMSV for all squeezing ranges. Especially, for the PS-then-PA TMSV $%
\left( s=0\right) $, it presents the largest APN compared with those for the PA-TMSV and the PS-TMSV. This means that our scheme can
show the advantage in terms of the total APN, which is beneficial for the improvement
of QFI. We also notice that, compared with the PA-TMSV, the APN of the GSP-TMSV when
 $s=0.5$ $(s=1)$ is smaller at $z<0.18$ ($%
z<0.4$).

\subsubsection{Antibunching effect of the GSP-TMSV}

In this subsection, let us consider the nonclassical properties of the
GSP-TMSV through the anti-bunching effect, which reflects the sub-Poisson
distribution implying the existence of nonclassical states \cite{48}. For an
arbitrary two-mode system, generally, the criteria of the antibunching
effect turns out to be \cite{20,49}%
\begin{equation}
R_{a,b}=\frac{\left \langle a^{\dagger 2}a^{2}\right \rangle +\left \langle
b^{\dagger 2}b^{2}\right \rangle }{2\left \langle a^{\dagger }ab^{\dagger
}b\right \rangle }-1.  \label{9}
\end{equation}

\begin{figure}[tbp]
\label{Fig4} \centering \includegraphics[width=0.8\columnwidth]{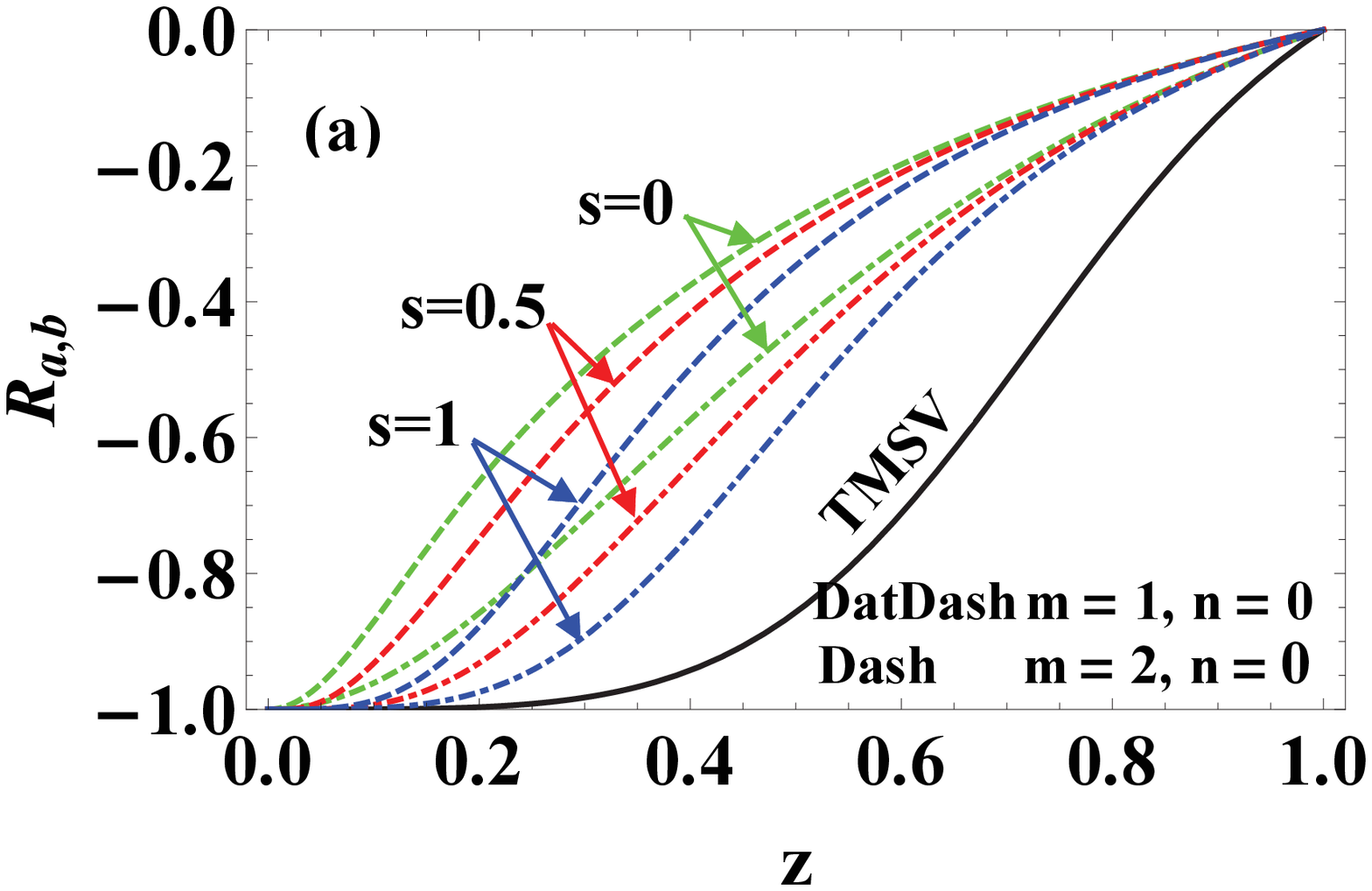} %
\includegraphics[width=0.8\columnwidth]{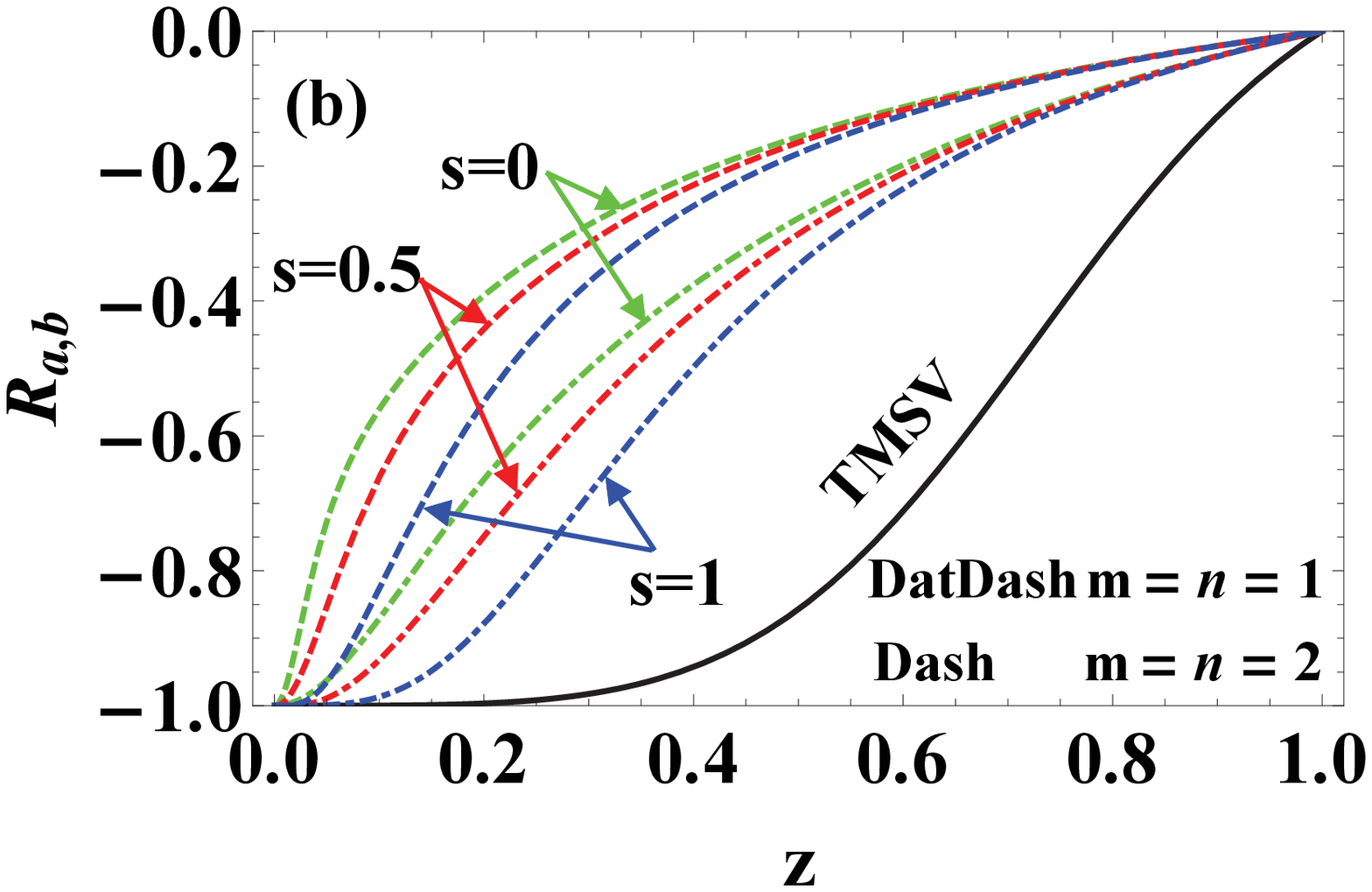}
\caption{{}(Color online) The antibunching effect $R_{a,b}$ as a function of
squeezing parameter $z$ for different operator parameter $s=0,0.5,1$. for
(a) the single-side GSP operations ($\left( m,n\right) \in \left \{ \left(
0,1\right) ,\left( 0,2\right) \right \} $), (b) the two-side symmetric GSP
operations ($\left( m,n\right) \in \left \{ \left( 1,1\right) ,\left(
2,2\right) \right \} $). Solid lines correspond to the TMSV case.}
\end{figure}

According to Eq. (\ref{6}), we can obtain the explicit expression of
anti-bunching effect $R_{a,b}$ in theory. In principle, the condition of $%
R_{a,b}<0$ corresponds to the existence of the antibunching effect, which
means that this quantum state has the nonclassicality. To clearly see this
point, in Fig. 4, we show the antibunching effect $R_{a,b}$ as the function
of squeezing parameter $z$ for different several superposition values $%
s=0,0.5,1$, together with the single-side ($\left( m,n\right) \in \{ \left(
0,1\right) ,(0,2)\}$) and the two-side symmetric GSP operations ($\left(
m,n\right) \in \{ \left( 1,1\right) ,(2,2)\}$). It is found that the
GSP-TMSV states, involving the single-side GSP case (see Fig. 4(a)) and the
two-side symmetric GSP cases (see Fig. 4(b)) always present the
anti-bunching effect, which indicates the usage of the GSP operations make
it possible to show the nonclassicality. However, this criteria of the
antibunching effect can not reflect how the change of $s=0,0.5,1$ in our
scheme affects the strength of the nonclassicality.

\subsubsection{Two-mode squeezing property}

To solve the aforementioned problem, in this subsection, we further
discusses the two-mode squeezing property of the GSP-TMSV state by using $%
\left \langle \Delta X_{1}^{2}\right \rangle $ and $\left \langle \Delta
X_{2}^{2}\right \rangle $, where $\left \langle \Delta
X_{i}^{2}\right
\rangle =\left \langle X_{i}^{2}\right \rangle
-\left
\langle X_{i}\right
\rangle ^{2}$ $\left( i=1,2\right) $ and $X_{1}$
$\left( X_{2}\right) $ are the sum (difference) of the orthogonal components
of $X_{a}$ and $X_{b},$ i.e. $X_{1}=X_{a}+X_{b}$ ($X_{2}=X_{a}-X_{b}$) with $%
X_{a}=\left( ae^{-i\theta _{1}}+a^{\dagger }e^{i\theta _{1}}\right) /\sqrt{2}
$ and $X_{b}=\left( be^{-i\theta _{2}}+a^{\dagger }e^{i\theta _{2}}\right) /%
\sqrt{2} $. For a given two-mode system, its two-mode variances are given by
\cite{50}%
\begin{equation}
\left \langle \Delta X_{1,2}^{2}\right \rangle =1+2\left \langle a^{\dagger
}a\right \rangle \pm 2\left \langle ab\right \rangle \cos (\theta
_{1}+\theta _{2}),  \label{10}
\end{equation}%
For simplicity, here we take $\theta _{1}+\theta _{2}=\pi .$ From Eqs. (\ref%
{6}) and (\ref{10}), when $m=n=0$, we can obtain $\left \langle \Delta
X_{1}^{2}\right \rangle =\left( 1-z\right) /\left( 1+z\right) $ and $%
\left
\langle \Delta X_{2}^{2}\right \rangle =\left( 1+z\right) /\left(
1-z\right) ,$ which are compatible with the TMSV case, as expected. Note
that, for the two-mode vacuum state $\left \vert 00\right \rangle ,$ $%
\left
\langle \Delta X_{1}^{2}\right \rangle |_{\left \vert
00\right
\rangle }=\left \langle \Delta X_{2}^{2}\right \rangle
|_{\left
\vert 00\right
\rangle }=1$, which is a standard noise.
Therefore, by using a logarithmic scale defined as dB$[X_{1}|_{\left \vert
\psi \right \rangle }]=10\log _{10}\left[ \left \langle \Delta
X_{1}^{2}\right \rangle |_{\left
\vert \psi \right \rangle }/\left
\langle
\Delta X_{1}^{2}\right
\rangle |_{\left
\vert 00\right \rangle }\right] $
and dB$[X_{2}|_{\left
\vert \psi \right
\rangle }]=10\log _{10}\left[
\left \langle \Delta X_{2}^{2}\right
\rangle |_{\left \vert \psi
\right
\rangle }/\left \langle \Delta X_{2}^{2}\right
\rangle
|_{\left
\vert 00\right \rangle }\right] ,$ one can quantify the two-mode
squeezing property of an arbitrary two-mode quantum state $\left
\vert \psi
\right
\rangle .$ If dB$[X_{1}|_{\left
\vert \psi \right \rangle }]<0$ or
dB$[X_{2}|_{\left \vert \psi \right
\rangle }]<0,$ in general, the state $%
\left
\vert \psi \right \rangle $ can be viewed as a squeezed state.
\begin{figure}[tbp]
\label{Fig5} \centering \includegraphics[width=0.8\columnwidth]{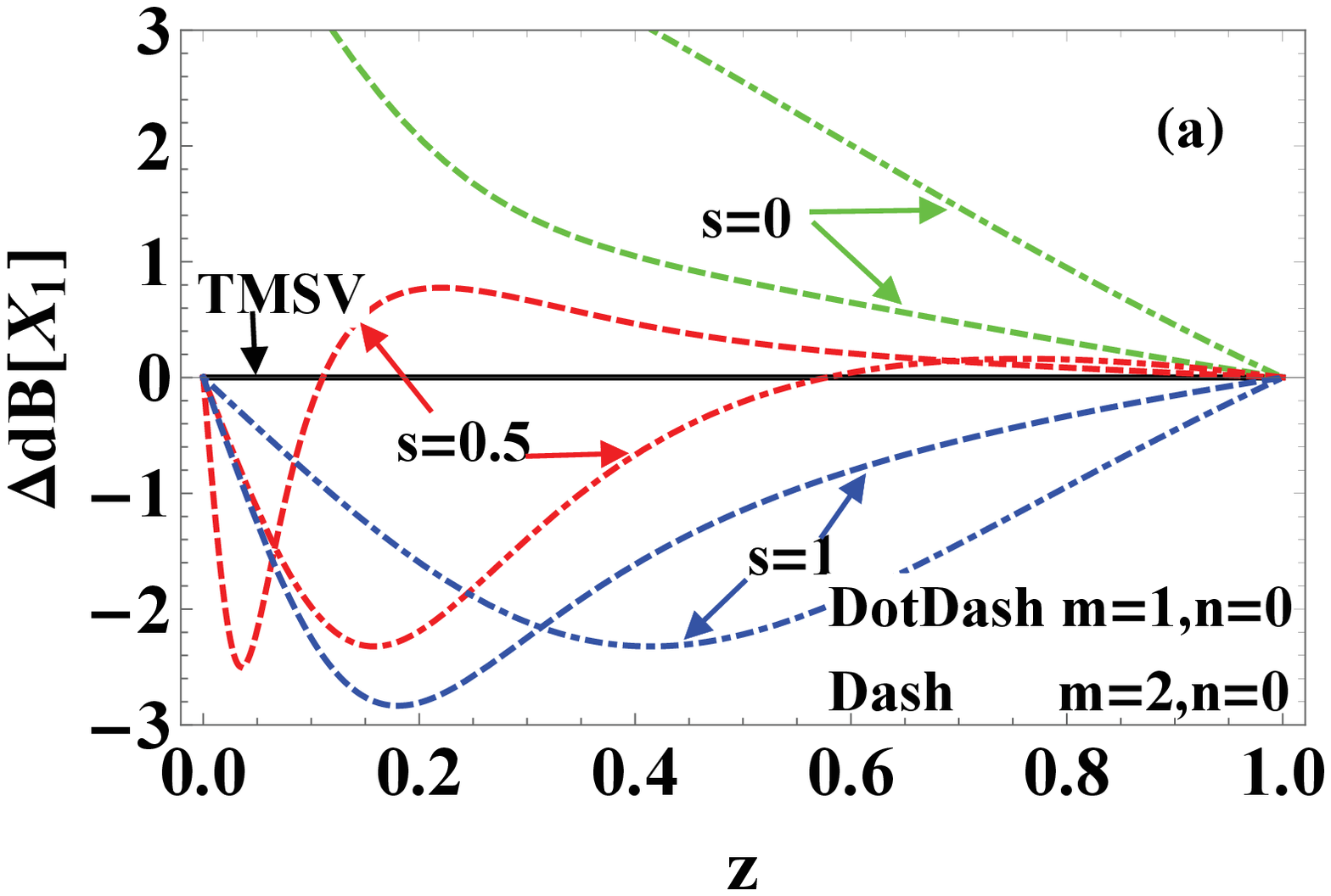} %
\includegraphics[width=0.8\columnwidth]{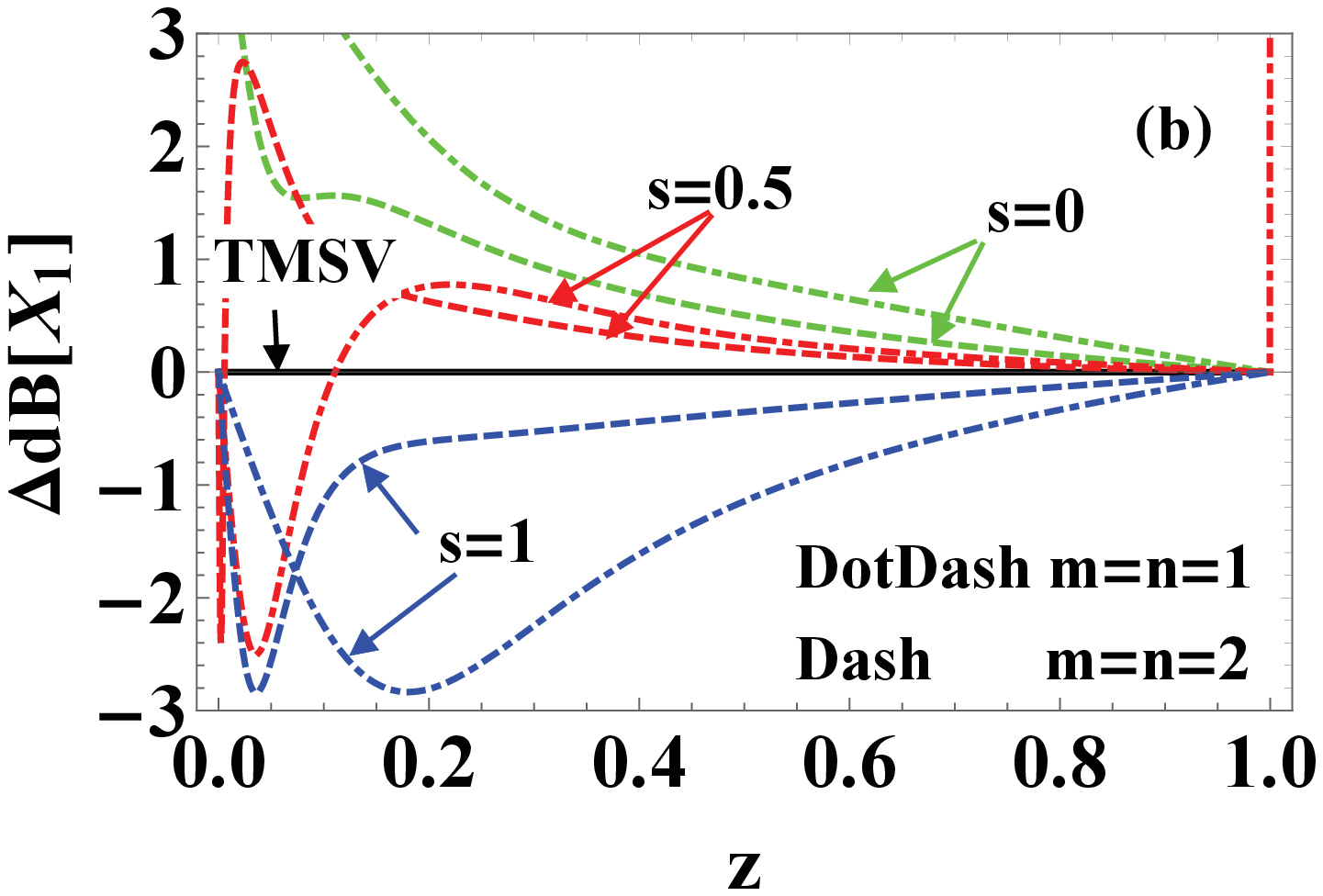}
\caption{{}(Color online) The two-mode squeezing property $%
dB[X_{1}|_{\left
\vert \protect \psi \right \rangle }]$ as a function of
squeezing parameter $z$ for different operator parameter $s=0,0.5,1$. for
(a) the single-side GSP operations ($\left( m,n\right) \in \left \{ \left(
0,1\right) ,\left( 0,2\right) \right \} $), (b) the two-side symmetric GSP
operations ($\left( m,n\right) \in \left \{ \left( 1,1\right) ,\left(
2,2\right) \right \} $). Solid lines correspond to the TMSV case.}
\end{figure}

To study the improvement of two-mode squeezing property between the GSP-TMSV
and the initial TMSV, in Fig. 5, we plot the difference $\Delta $dB$[X_{1}]=
$ $10\log _{10}\left[ \left \langle \Delta X_{1}^{2}\right \rangle |_{\left
\vert \psi \right \rangle _{ab}}/\left \langle \Delta X_{1}^{2}\right
\rangle |_{TMSV}\right] $ as the function of $z$ with several superposition
values $s=0,0.5,1$, including the single-side GSP operations ($\left(
m,n\right) \in \{ \left( 0,1\right) ,(0,2)\}$) and the two-side symmetric
GSP operations ($\left( m,n\right) \in \{ \left( 1,1\right) ,(2,2)\} $). In
principle, the condition of $\Delta $dB$[X_{1}]<0$ means the existence and
improvement of two-mode squeezing property, but $\Delta $dB$[X_{1}]\geqslant
0$ only indicates that two-mode squeezing property cannot be enhanced. It is
interesting that, as for the two types of the GSP operations, the improved
area of two-mode squeezing property for $s=0 $ can not be shown, which means
that using PS-then-PA operation on the TMSV makes it impossible to present
the improvement of two-mode squeezing property. Whereas for other cases for $%
s=0.5$ and $1,$ the latter can always show the existence and improvement of
two-mode squeezing property, and the improved area of two-mode squeezing
property for the former would be limited at a small squeezing range.
Besides, with the increase of $\left( m,n\right) $, this limitation is more
obvious with respect to the narrower of achievable squeezing ranges. We also
notice that, at a fixed $s,$ for the case of $s=0.5 $, the achievable
squeezing range for the single-side GSP operations are bigger than that for
the two-side symmetric GSP operations in terms of the improvement of the
two-mode squeezing property.

\section{Improvement of the QFI via the GSP-TMSV}

After evaluating the nonclassical properties of the GSP-TMSV, then we
consider whether the GSP-TMSV can be used to improve the QFI when the
GSP-TMSV is used as inputs of the balanced MZI, which consists of two
symmetrical beam splitters (BSs), shown in Fig. 1 (Box 2). In Ref. \cite{51}%
, it is pointed out that the behavior of a BS can be described as a
rotation, i.e., using the Schwinger representation of SU(2) algebra,
\begin{eqnarray}
J_{1} &=&\frac{1}{2}\left( a^{\dagger }b+ab^{\dagger }\right) ,\text{ }J_{2}=%
\frac{1}{2i}\left( a^{\dagger }b-ab^{\dagger }\right) ,  \notag \\
J_{3} &=&\frac{1}{2}\left( a^{\dagger }a-b^{\dagger }b\right) ,\text{ }J_{0}=%
\frac{1}{2}\left( a^{\dagger }a+b^{\dagger }b\right) ,  \label{11}
\end{eqnarray}%
where $J_{0}$ is a Casimir operator that commutes with all others angular
momentum operators $\left[ J_{i},J_{0}\right] =0$ $\left( i=1,2,3\right) $,
which should satisfy the commutation relation $\left[ J_{i},J_{j}\right]
=i\varepsilon _{ijk}J_{k}$ $\left( i,j,k=1,2,3\right) $, then the action of
the MZI can be equivalent to the following unitary operator%
\begin{equation}
U\left( \varphi \right) =e^{i\pi J_{1}/2}e^{-i\varphi J_{3}}e^{-i\pi
J_{1}/2}=e^{-i\varphi J_{2}}.  \label{12}
\end{equation}%
Thus, when inputting any pure state $\left \vert in\right \rangle $ into the
MZI, the output state is given by%
\begin{equation}
\left \vert out\right \rangle _{MZI}=e^{-i\varphi J_{2}}\left \vert in\right
\rangle .  \label{13}
\end{equation}%
Combining Eqs. (\ref{2}) and (\ref{13}), for our scheme, the resulting state
prior to the parity detection can be derived as
\begin{equation}
\left \vert out\right \rangle _{MZI}=\frac{\Re u}{\sqrt{P_{d}}}e^{a^{\dagger
}b^{\dagger }v\cos \varphi +\frac{1}{2}(b^{\dagger 2}-a^{\dagger 2})v\sin
\varphi }\left \vert 00\right \rangle ,  \label{14}
\end{equation}%
where we have used $e^{i\varphi J_{2}}\left \vert 00\right \rangle
=\left
\vert 00\right \rangle $ and the following transformation relations,
\begin{eqnarray}
e^{-i\varphi J_{2}}a^{\dagger }e^{i\varphi J_{2}} &=&a^{\dagger }\cos \frac{%
\varphi }{2}+b^{\dagger }\sin \frac{\varphi }{2},  \notag \\
e^{-i\varphi J_{2}}b^{\dagger }e^{i\varphi J_{2}} &=&b^{\dagger }\cos \frac{%
\varphi }{2}-a^{\dagger }\sin \frac{\varphi }{2}.  \label{15}
\end{eqnarray}%
In particular, for the case of $m=n=0$, Eq. (\ref{14}) reduces to
\begin{equation}
\left \vert TMSV\right \rangle \text{=}\sqrt{1-z^{2}}e^{z[a^{\dagger
}b^{\dagger }\cos \varphi +\frac{1}{2}(b^{\dagger 2}-a^{\dagger 2})\sin
\varphi ]}\left \vert 00\right \rangle ,  \label{16}
\end{equation}%
which is just the result in Ref. \cite{8}, where the TMSV is used as inputs
of the MZI, and the superresolution and sub-Heisenberg sensitivity can be
achieved using parity detection. It is interesting that, due to the fact
that the usefulness of non-Gaussian (PA- and PS-) operations for achieving
the strongly nonclassical states, the PS(PA-)-based TMSV scheme has been
proposed for further improving the measurement precision of quantum
metrology. Then a question naturally arises: can our proposed GSP-TMSV
scheme improve the phase sensitivity and resolution in quantum metrology?

Next, we first consider the proposed GSP-TMSV as the input of the MZI to
study its QFI denoted by $F_{Q}$. The QFI is associated with the ultimate
limit of phase sensitivity, which is given by the quantum Cramer-Rao
boundary (QCRB) \cite{52}, i.e.,
\begin{equation}
\Delta \phi _{\min }=\frac{1}{\sqrt{F_{Q}}}.  \label{17}
\end{equation}%
In particular, for any pure state $\left \vert \psi \left( \theta \right)
\right \rangle ,$ the QFI can be calculated as
\begin{equation}
F_{Q}=4\left \{ \left \langle \psi ^{^{\prime }}\left( \theta \right) \right
\vert \left. \psi ^{^{\prime }}\left( \theta \right) \right \rangle -\left
\vert \left \langle \psi ^{^{\prime }}\left( \theta \right) \right. \left
\vert \psi \left( \theta \right) \right \rangle \right \vert ^{2}\right \} ,
\label{18}
\end{equation}%
where $\left \vert \psi \left( \theta \right) \right \rangle =e^{-i\theta
J_{3}}e^{-i\pi J_{1}/2}\left \vert in\right \rangle $ and $\left \vert \psi
^{^{\prime }}\left( \theta \right) \right \rangle =\partial \left \vert \psi
\left( \theta \right) \right \rangle /\partial \theta $. Thus, for the
GSP-TMSV state shown in Eq. (\ref{2}), the QFI can be directly calculated as
\begin{equation}
F_{Q}=2N\left( N+1\right) -\left \langle in\right \vert \left( a^{\dagger
2}b^{2}+a^{2}b^{\dagger 2}\right) \left \vert in\right \rangle ,  \label{19}
\end{equation}%
where the APN $N$ has the same definition as Eq. (\ref{8}) and the second
term can be derived using Eq. (\ref{6}). Especially, for the case of $m=n=0$
corresponding to the TMSV as inputs, Eq. (\ref{19}) reduces to $F_{Q}$ $%
=4z^{2}/\left( 1-z^{2}\right) ^{2}$, as expected \cite{22}.

According to Eq. (\ref{19}), we illustrate the QFI as a function of $z$ for
the single-side ($\left( m,n\right) \in \{ \left( 0,1\right) ,(0,2)\}$) and
the two-side symmetric GSP operations ($\left( m,n\right) \in \{ \left(
1,1\right) ,(2,2)\}$), as shown in Figs. 6(a) and 6(b), respectively. It is
obvious that the QFI using TMSV input (the black solid line) is outperformed
by that using the GSP-TMSV for these two cases above. Specifically speaking,
when given some parameters $s$ and $z$, the QFI of our scheme increases with
the increase of $\left( m,n\right) $, especially for two-side symmetric GSP
operations. The reason may be the fact that the APN of the GSP-TMSV
increases as the increasing $\left( m,n\right) $ (see Fig. 2). In addition,
at some fixed parameters $\left( m,n\right) $ and $z$, it is found that the
QFI corresponding to the PS-then-PA operation ($s=0$) is always better than
other cases, including $s=1$ and $s=0.5$. In addition, compared to the cases
with $s=0$ and $s=0.5$, the QFI using PA-then-PS operation has a relatively
poor improvement.

In order to highlight the advantages of the GSP-TMSV as the input of the
MZI, we further make a comparison about the QFI for several different
non-Gaussian states, such as single PA-TMSV (magenta dashed), single PS-TMSV
(cayan dashed) and the GSP-TMSV with $m=n=1$. The QFI as a function of
squeezing parameter $z$ is plotted in Fig. 7. It is interesting that both PA
and PS operations always achieve an improvement of the QFI compared to the
TMSV in the whole squeezing parameter region, while the PA operation
presents a better performance than the PS operation. In addition, for the
two cases with $s=1$ and $s=0.5$, the QFI can be also improved when the
squeezing parameter exceeds a small threshold. The latter with $s=0.5$
performs better than the former with $s=1.$ However, among these
non-Gaussian operations, the PS-then-PA operation ($s=0$) presents the best
improvement in the whole squeezing parameter region. These results are
similar to the APN cases of different (non-)Gaussian states (see Fig.3).

\begin{figure}[tbp]
\label{Fig6} \centering \includegraphics[width=0.8\columnwidth]{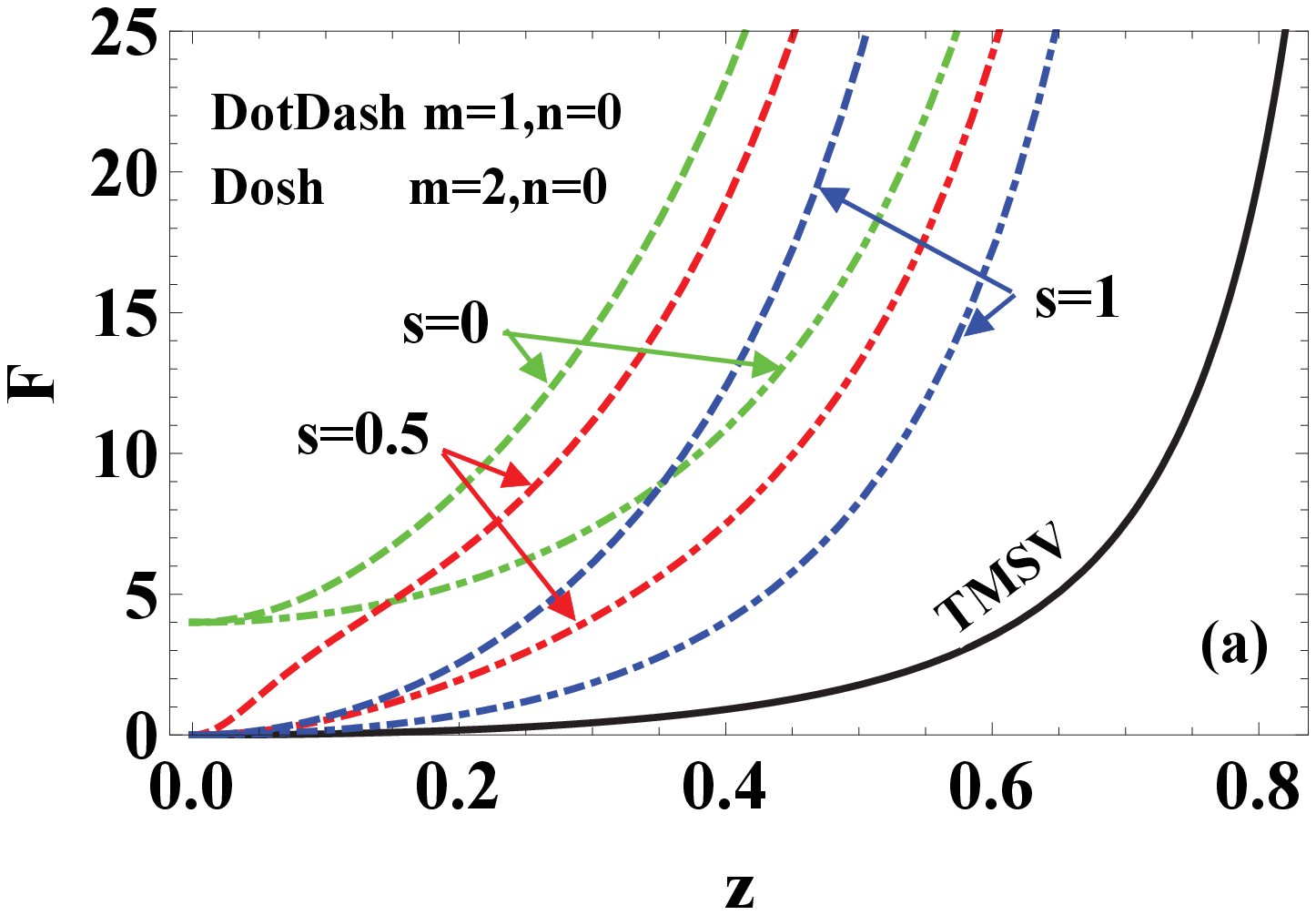} %
\includegraphics[width=0.8\columnwidth]{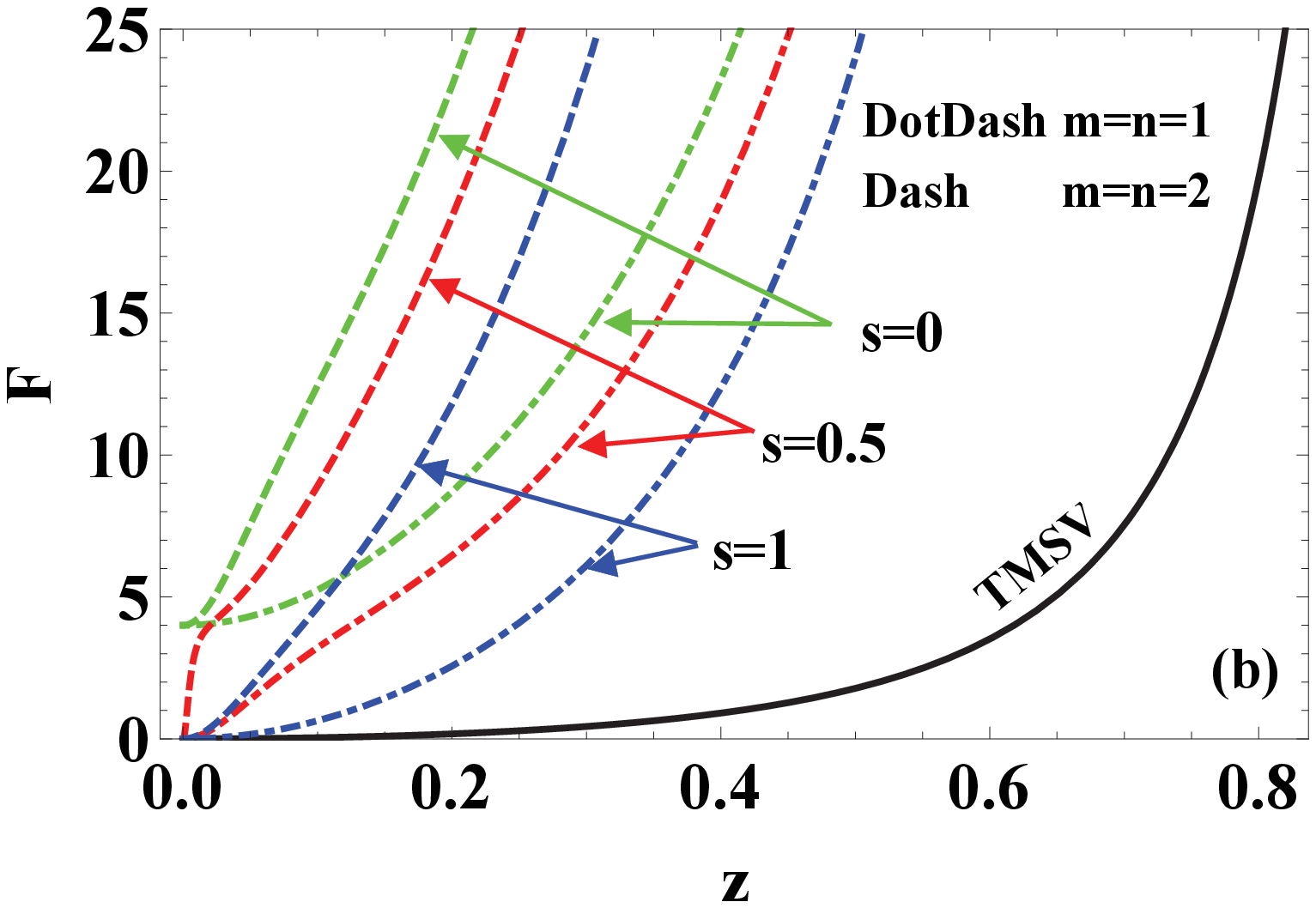}
\caption{{}(Color online) Plots of the quantum Fisher information $F_{Q}$
against the squeezing parameter $z$ for different operator parameter $%
s=0,0.5,1$. for (a) the single-side GSP operations ($\left( m,n\right) \in
\left \{ \left( 0,1\right) ,\left( 0,2\right) \right \} $), (b) the two-side
symmetric GSP operations ($\left( m,n\right) \in \left \{ \left( 1,1\right)
,\left( 2,2\right) \right \} $). Solid lines correspond to the TMSV case.}
\end{figure}

\begin{figure}[tbp]
\label{Fig7} \centering \includegraphics[width=7.2cm]{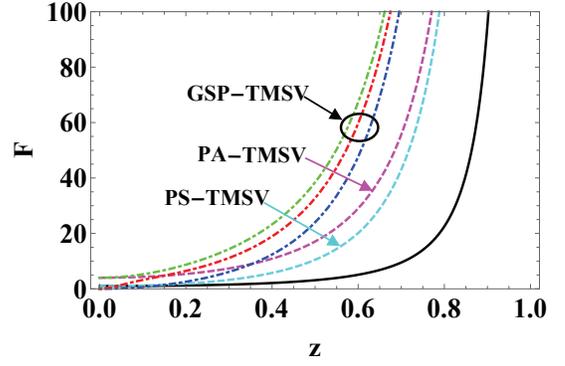}
\caption{{}(Color online) As a comparison, the QFI $F_{Q}$ as a function of
the squeezing parameter $z$. The dot-dashed lines represent our scheme for
operation parameter $s=0,0.5,1$ (corresponding to green, red, and blue color
line, respectively.), and dashed lines represent the previous work of
performing the PA-TMSV (magenta color line) and the PS-TMSV (cyan color
line). Solid line corresponds to the TMSV case. }
\end{figure}

\section{Phase estimation with parity detection}

In this section, we considered the QFI, which corresponds to the upper bound
of measurement preision. Actually, the practical precision depends on the
way of measure. In this section, we further examine the phase estimation
using special measures. Note that the parity detection has advantages over
the other detection schemes, thus here we shall take the parity detection as
a powerful tool for analyzing the phase sensitivity of our scheme.

\subsection{The parity detection}

In fact, the aim of parity detection is to obtain the expectation value of
the parity operator in the output state of the MZI \cite{53}, which plays a
vital role in quantum measurements. In particular, when the TMSV is used as
the input, the parity detection can effectively extract the phase
information, while the intensity detection is not applicable \cite{45}. For
convenience, we choose the $b$ mode of the output, then the parity operator
can be written as%
\begin{equation}
\Pi _{b}=e^{i\pi b^{\dagger }b}=\int \frac{d^{2}\gamma }{\pi }\left \vert
\gamma \right \rangle \left \langle -\gamma \right \vert ,  \label{20}
\end{equation}%
where $\left \vert \gamma \right \rangle $ is the coherent state, such that
for an arbitrary output state $\rho _{out}=\left \vert out\right \rangle
_{MZI}\left \langle out\right \vert $ in the MZI, the corresponding
expectation value of $\Pi _{b}$ can be expressed as
\begin{equation}
\left \langle \Pi _{b}\right \rangle =\mathtt{Tr}[\Pi _{b}\rho _{out}]=\int
\frac{d^{2}\gamma }{\pi }\left \langle -\gamma \right \vert \rho _{out}\left
\vert \gamma \right \rangle .  \label{21}
\end{equation}%
Thus, based on Eq. (\ref{13}), the expectation value $\left \langle \Pi
_{b}\right \rangle $ can be calculated as

\begin{equation}
\left \langle \Pi _{b}(\varphi )\right \rangle =\widetilde{\Re }\frac{%
uu_{1}\Omega _{1}}{P_{d}\sqrt{\Omega _{2}-\Omega _{3}}},  \label{22}
\end{equation}%
with%
\begin{eqnarray}
\Omega _{1} &=&\left( 1-v_{1}v\sin ^{2}\varphi \right) ^{\frac{1}{2}},
\notag \\
\Omega _{2} &=&\left( v_{1}v\cos 2\varphi +1\right) ^{2},  \notag \\
\Omega _{3} &=&v_{1}v\left( vv_{1}-1\right) ^{2}\sin ^{2}\varphi .
\label{23}
\end{eqnarray}%
In particular, when $m=n=0,$ Eq. (\ref{22}) reduces to $\left \langle \Pi
_{b}(\varphi )\right \rangle =\left( 1-z^{2}\right) /\sqrt{\left(
1-2z^{2}\cos 2\phi +z^{4}\right) }$ ($\varphi =\phi +\pi /2$)$,$
corresponding to the TMSV case, as expected \cite{8}. In the following, we
will use the variable $\phi $ to investigate the resolution and sensitivity.

In Ref. \cite{8}, it has been shown that the central peak of $\left \langle
\Pi _{b}(\phi +\pi /2)\right \rangle $ at $\phi =0$ for the TMSV inputs is
narrower than that for the coherent state input under the same parameters,
thereby achieving superresolution and sub-Heisenberg sensitivity of the MZI.
However, it is interesting that the case can be further improved using our
scheme. For given squeezing parameter $z=0.6,$ using Eq. (\ref{22}) we
illustrate the expectation values $\left \langle \Pi _{b}(\phi +\pi
/2)\right \rangle $ as a function of the phase shift $\phi $ in Fig. 8,
including both the single-side GSP operations ($\left( m,n\right) \in \{
\left( 0,1\right) ,(0,2)\}$ in Fig. 8(a)) and the two-side symmetric GSP
operations ($\left( m,n\right) \in \left \{ \left( 1,1\right) ,\left(
2,2\right) \right \} $ in Fig. 8(b)).

From Fig. 8, it is clear that the central peak of $\left \langle \Pi
_{b}(\phi +\pi /2)\right \rangle $ at $\phi =0$ for all the GPS-TMSV inputs
is much narrower than that for the TMSV input. It implies that the use of
the GSP operation is beneficial for significantly increasing the
superresolution. Among these non-Gaussian operations, the PS-then-PA
operation ($s=0$) presents the best performance again. In addition, for both
the single-side (Fig. 8(a)) and two-side (Fig. 8(b)) GSP operations, the
resolution can be further enhanced by increasing the parameter $\left(
m,n\right) $. Compared to the single-side case, the two-side case has a
better performance for the improvement of superresolution under the same
parameters.

In Fig. 9, we make a comparison about $\left \langle \Pi _{b}(\phi +\pi
/2)\right \rangle $ between single PA(PS)-TMSVs and our proposed scheme with
$m=n=1$ for a given squeezing parameter $z=0.6$. It is obvious that these
non-Gaussian operations can effectively enhance the resolution and the
effects of improvement can be ranked from small to large, i.e., PS, PA,
PA-then-PS ($s=1$), PA-then-PS plus PS-then-PA ($s=0.5$), and PS-then-PA ($%
s=0$). Thus, compared to both PA and PS, our scheme presents the advantages
for further improving superresolution, especially for PS-then-PA $(s=0).$%
\begin{figure}[tbp]
\label{Fig8} \centering \includegraphics[width=1\columnwidth]{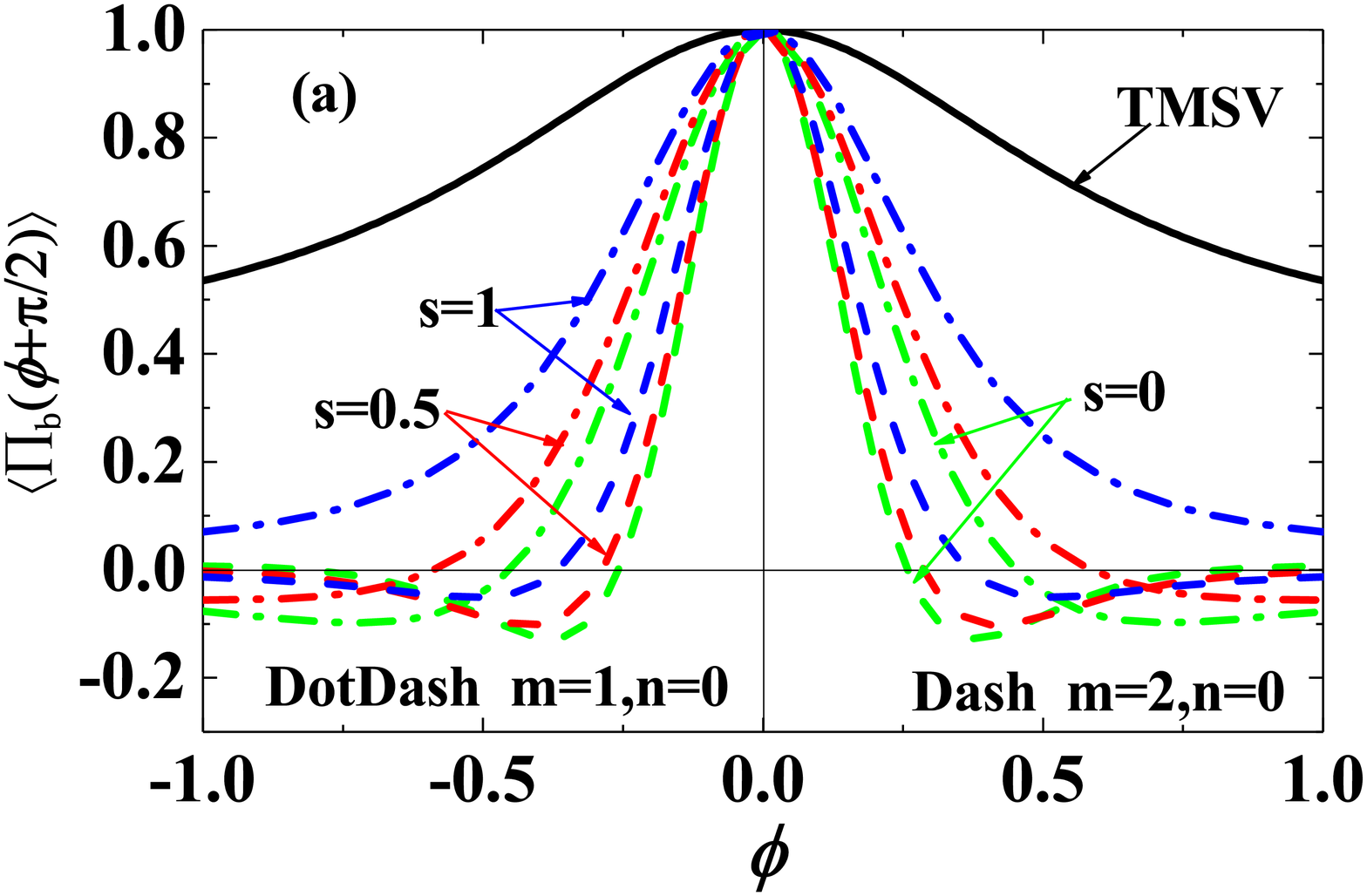} %
\includegraphics[width=1\columnwidth]{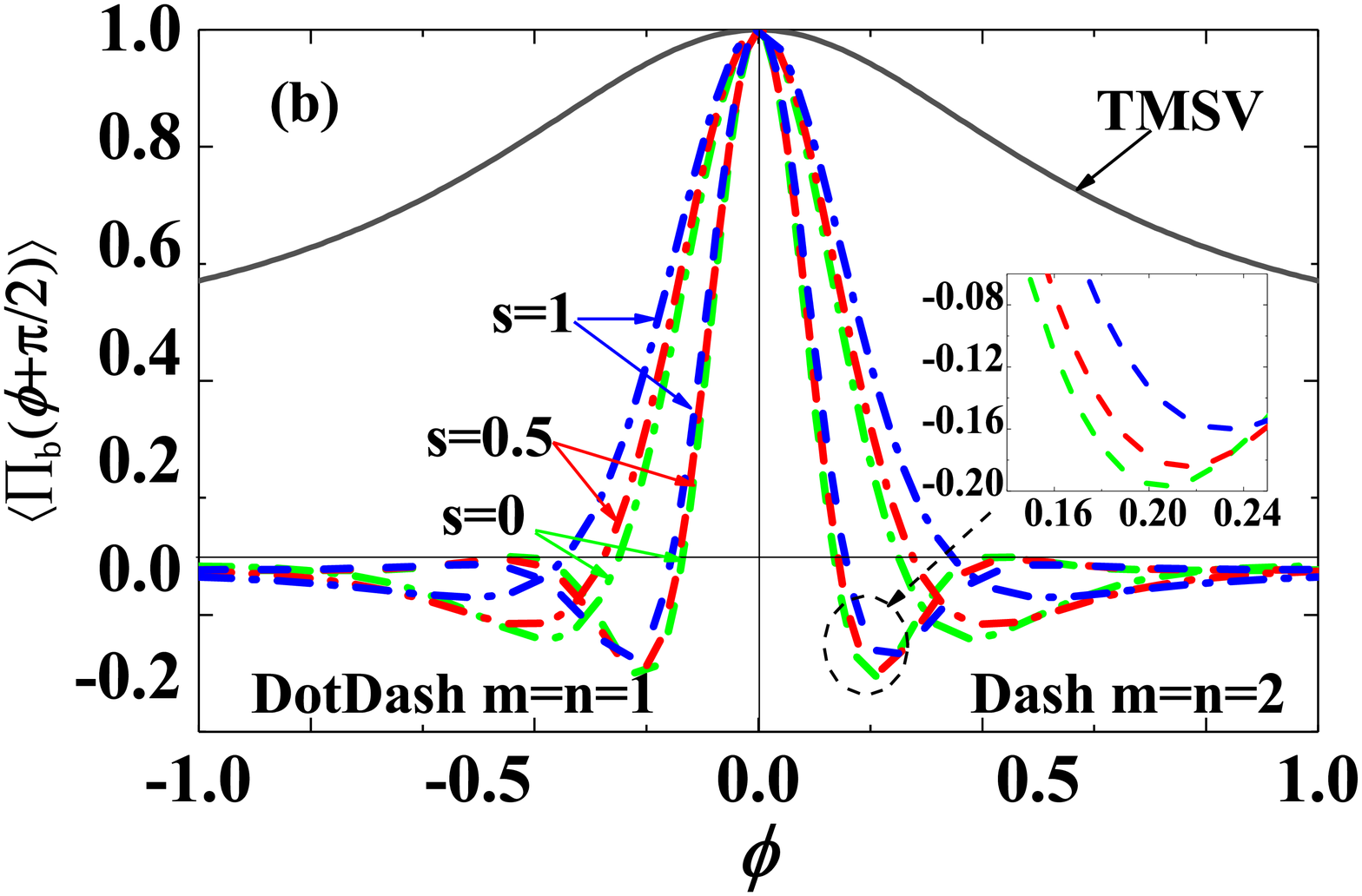}
\caption{{}(Color online) The expectation values of the parity operator $%
\left \langle \Pi _{b}(\protect \phi +\protect \pi /2)\right \rangle $ versus
the phase shift $\protect \phi $ for a given squeezed parameter $z=0.6$ and
different operator parameter $s=0,0.5,1$. (a) the single-side GSP operations
($\left( m,n\right) \in \left \{ \left( 0,1\right) ,\left( 0,2\right)
\right
\} $), (b) the two-side symmetric GSP operations ($\left( m,n\right)
\in \left \{ \left( 1,1\right) ,\left( 2,2\right) \right \} $). Solid lines
correspond to the TMSV case.}
\end{figure}
\begin{figure}[tbp]
\label{Fig9} \centering \includegraphics[width=7.2cm]{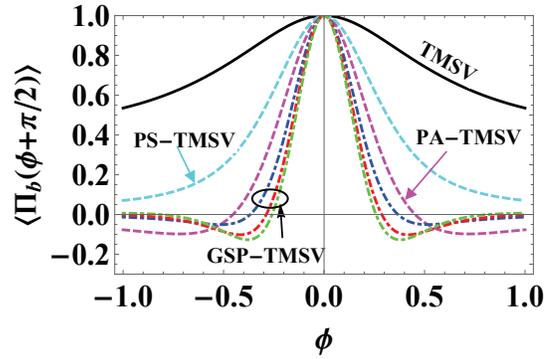}
\caption{{}(Color online) The expectation values of the parity operator $%
\left \langle \Pi _{b}(\protect \phi +\protect \pi /2)\right \rangle $ as a
function of $\protect \phi $ for fixed squeezed parameter $z=0.6$ for
different non-Gaussian operations. The dot-dashed lines represent the our
work for operation parameter $s=0,0.5,1$ (green, red, and blue color line,
respectively.), and dashed lines represent the previous work performing the
PA (magenta color line) and the PS operations (cyan color line). Solid line
corresponds to the TMSV case. }
\end{figure}

\subsection{The phase sensitivity}

After investigating the resolution of our scheme in the MZI, in this
subsection, we further consider the sensitivity of phase estimation based on
the outcome of parity detection. In general, the phase sensitivity of the
MZI can be estimated by the error propagation formula \cite{54,55}, i.e.,
\begin{equation}
\Delta \phi =\frac{\sqrt{1-\left \langle \Pi _{b}(\varphi )\right \rangle
^{2}}}{|\partial \Pi _{b}/\partial \phi |}.  \label{24}
\end{equation}%
In particular, when $m=n=0$ corresponding to the case of TMSV input, using
Eq. (\ref{22}) then Eq. (\ref{24}) reduces to $\Delta \phi _{TMSV}=\left(
1-2z^{2}\cos 2\phi +z^{4}\right) /\left \{ [2z(1-z^{2})\cos \phi ]\right \}
, $ as expected. At the limitation of $\phi \rightarrow 0,$ $\Delta \phi
_{TMSV}$ becomes $\Delta \phi _{\min }=\left( 1-z^{2}\right) /\left(
2z\right) =1/\sqrt{F_{Q}}$, in which $F_{Q}$ is the QFI for the TMSV input
into the MZI. This indicates that the QCRB can be achieved especially at $%
\phi \rightarrow 0$ with the help of the parity detection.
\begin{figure}[tbp]
\label{Fig10} \centering \includegraphics[width=1\columnwidth]{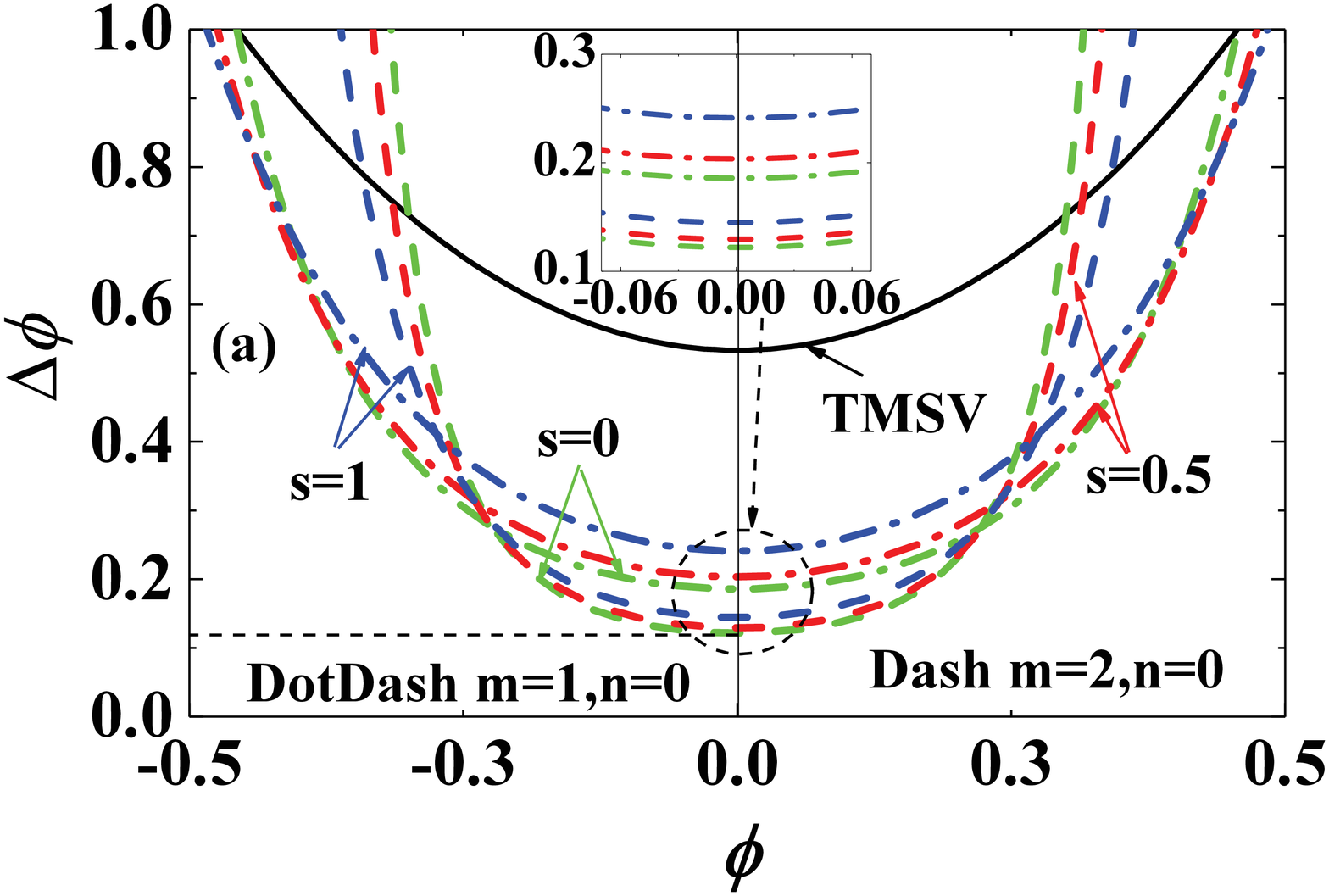} %
\includegraphics[width=1\columnwidth]{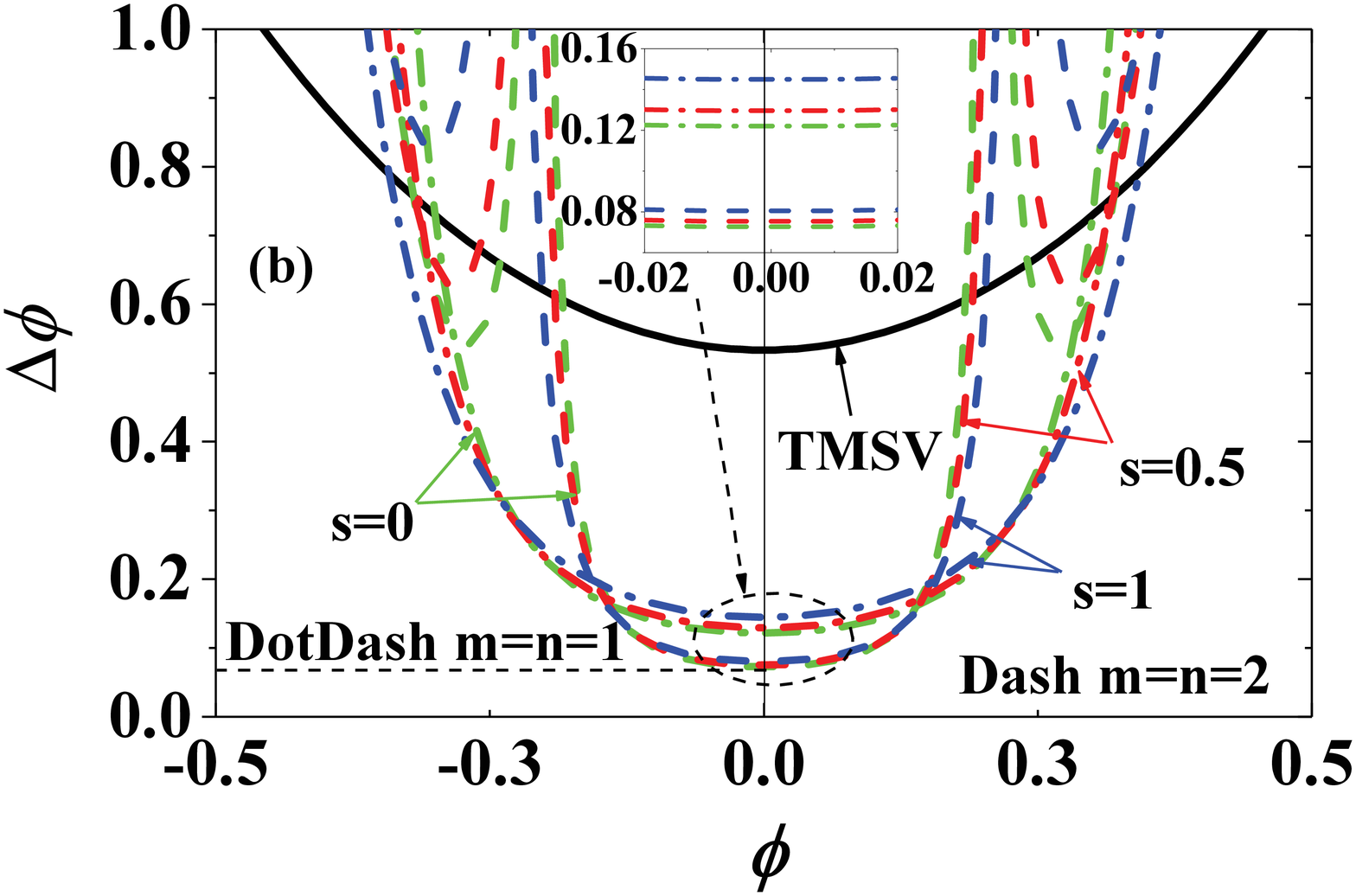}
\caption{{}(Color online) The phase uncertainty $\Delta \protect \phi $
versus the phase shift $\protect \phi $ for a given squeezed parameter $z=0.6$
and different operator parameter $s=0,0.5,1$. (a) the single-side GSP
operations ($\left( m,n\right) \in \left \{ \left( 0,1\right) ,\left(
0,2\right) \right \} $), (b) the two-side symmetric GSP operations ($\left(
m,n\right) \in \left \{ \left( 1,1\right) ,\left( 2,2\right) \right \} $).
Solid lines correspond to the TMSV case.}
\end{figure}

Generally, the lower the value $\Delta \phi $, the higher the phase
sensitivity. In order to clearly see the effects of different parameters on
the phase sensitivity, at fixed values of $z=0.6$ and $s=0,0.5,1$, we plot
the phase sensitivity$\  \Delta \phi $ as a function of the phase $\phi $ for
the single-side ($\left( m,n\right) \in \{ \left( 0,1\right) ,(0,2)\}$) and
the two-side ($\left( m,n\right) \in \left \{ \left( 1,1\right) ,\left(
2,2\right) \right \} $) symmetric GSP operations in Fig. 10(a) and 10(b),
respectively. Compared to the TMSV case, the minimum value of $\Delta \phi
_{\min }$ can be significantly reduced by single- and two-side cases above.
For given parameter $s$, the $\Delta \phi _{\min }$ can be further decreased
with the increasing of the parameters $\left( m,n\right) $. Comparing
single-side case with two-side case (Fig. 10(a) and 10(b)), it is clear that
the latter can achieve a lower $\Delta \phi _{\min }$ than the former. In
addition, the PS-then-PA ($s=0$) is the best operation for getting the
minimum value $\Delta \phi $ under the condition that other parameters are
the same.

In Fig. 11, we further make a comparison about $\Delta \phi _{\min }$
between single PA(PS)-TMSVs and our proposed scheme, where the condition is
the same as that in Fig. 10. In terms of minima $\Delta \phi _{\min },$ the
effects for these non-Gaussian operations can be ranked from large to small,
i.e., PS, PA, PA-then-PS ($s=1$), PA-then-PS plus PS-then-PA ($s=0.5$), and
PS-then-PA ($s=0$). Again, the PS-then-PA is the best choice for achieving
the minima of $\Delta \phi _{\min }$ due to the fact that the APN can be
increased by the PS-then-PA. These results indicate that under the same
parameters the phase sensitivity $\Delta \phi $ can be further enhanced by
using our scheme when comparing to the PA-TMSV and the PS-TMSV.

\begin{figure}[tbp]
\label{Fig11} \centering \includegraphics[width=8cm]{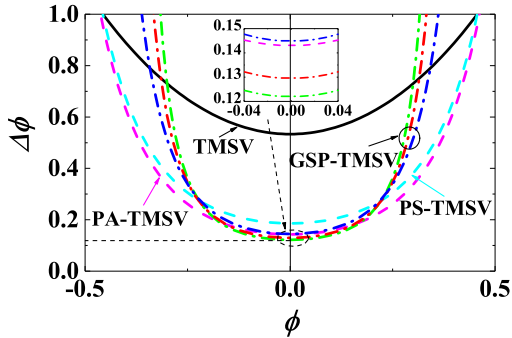}
\caption{{}(Color online) As a comparison, the phase uncertainty $\Delta
\protect \phi $ as a function of the phase shift $\protect \phi $ for fixed
squeezed parameter $z=0.6.$ The dot-dashed lines represent the our work for
operation parameter $s=0,0.5,1$ (corresponding to green, red, and blue color
line, respectively.), and dashed lines represent the previous work of
performing the PA-TMSV (magenta color line) and the PS-TMSV (cyan color
line). Solid line corresponds to the TMSV case}
\end{figure}

\begin{figure}[tbp]
\label{Fig12} \centering \includegraphics[width=0.8\columnwidth]{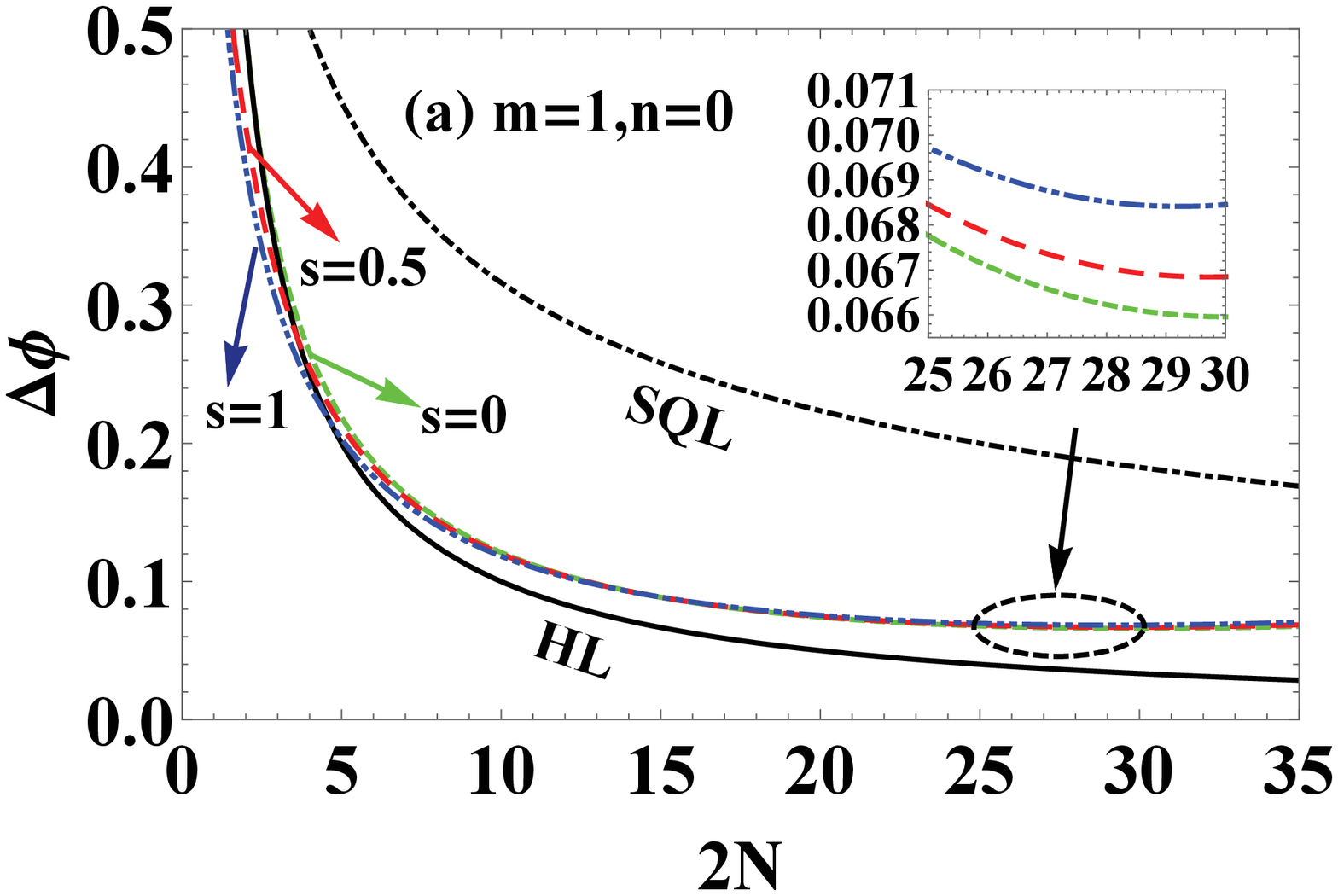} %
\includegraphics[width=0.8\columnwidth]{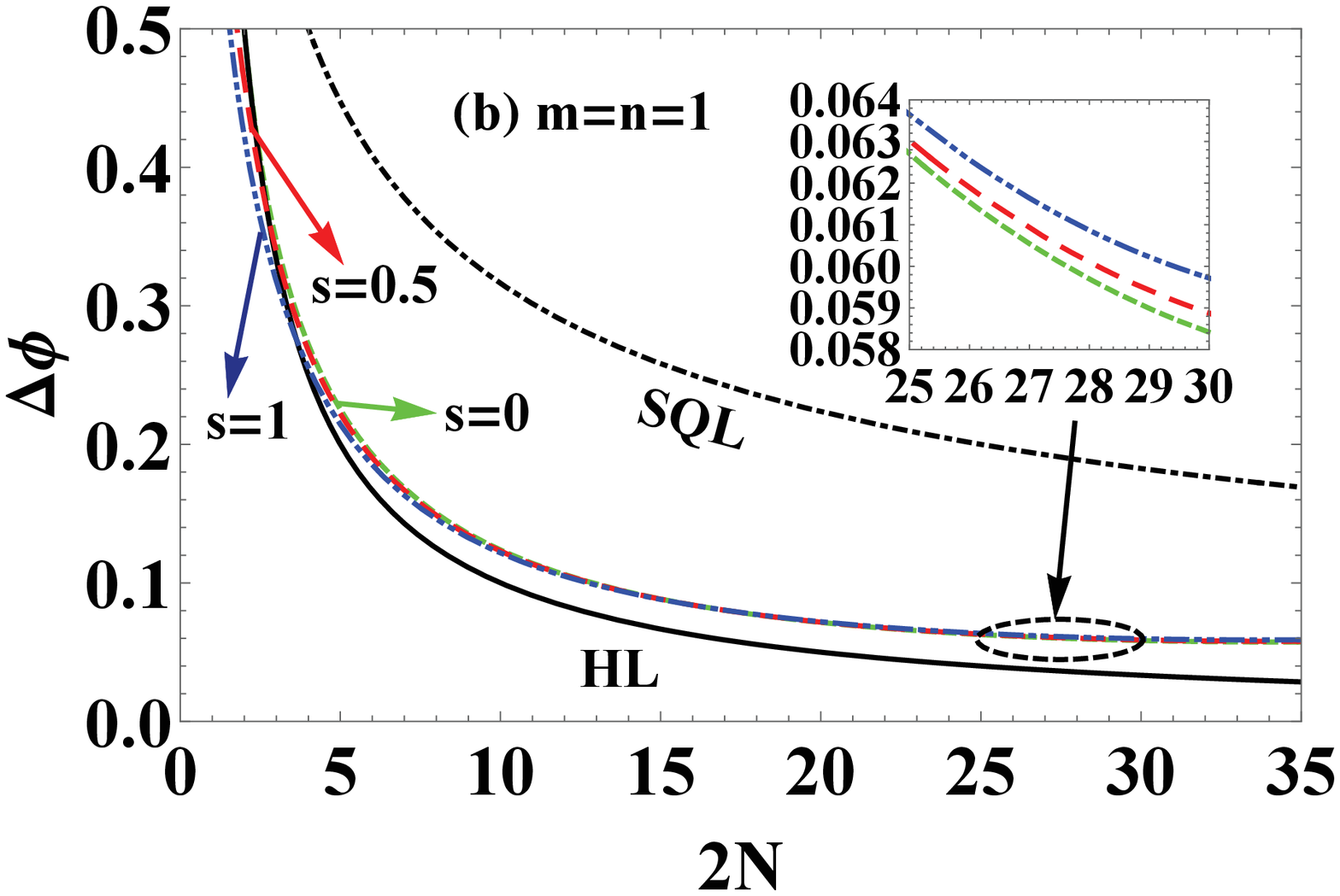}
\caption{{}(Color online) The phase sensitivity$\  \Delta \protect \phi $ as a
function of the total APN $2N$ for different operator parameter $s=0,0.5,1$.
for (a) the single-side GSP operations ($\left( m,n\right) \in \left(
0,1\right) $), (b) the two-side symmetric GSP operations ($\left( m,n\right)
\in \left( 1,1\right) $). Solid lines correspond to the TMSV case.}
\end{figure}

On the other hand, it is interesting to notice that the HL based on parity
detection can be beaten when the TMSV is considered as the input of the MZI
\cite{8}. However, when the PA-TMSV or PS-TMSV is used as inputs, the HL
cannot be beaten and the corresponding phase uncertainties perform worse
compared to the TMSV under the same parameters \cite{22}. Then how about our
scheme? In order to clearly see this point, for given phase $\phi =0.05$ and
$s=0,0.5,1$, we show the phase sensitivity as function of the total APN $2N$
for single-side ($\left( m,n\right) \in \left( 1,0\right) $) and two-side
symmetric GSP operations ($\left( m,n\right) \in \left( 1,1\right) $) in
Figs. 12(a) and 12(b), respectively. For our scheme, it is shown that the
SQL is always broken through due to the fact that the GSP-TMSV is a kind of
nonclassical state. As discussed above, the PS-then-PA ($s=0$) can be used
to achieve the best phase sensitivity and superresolution, however, the HL
cannot be beaten by this case, but by the cases of $s=0.5,1$ in the regime
of the small total APN (or say, the small initial squeezing parameter $z$).
The reason may be that, except for $s=0,$ the two-mode squeezing property
can be always improved for the cases of $s=0.5,1$ at the certain range of $%
z, $ which can be seen from Fig. 5. In addition, it is also interesting to
notice that, for the cases of $s=0.5,1$, it is much significant for beating
the HL using single-side GSP operation $\left( m,n\right) \in \left(
1,0\right) $ rather than two-side GSP operation $\left( m,n\right) \in
\left( 1,1\right) $ at small range of the total APN. While in the larger
total APN region, two-side GSP operation is much easier to make the phase
uncertainty close to the HL, which is beneficial for the practical
implementation of achieving the super-sensitivity.

\section{Effects of photon losses on phase sensitivity}

In practice, the travelling states are inevitably coupled to the
environment, so that the decoherence process should be taken into account.
Generally, there are several models of decoherence processes, such as photon
loss, phase diffusion and thermal noise. As described in Ref. \cite{56},
particularly, it is shown that the photon losses have a significant impact
on phase sensitivity. Thus, here we only consider the effects of photon loss
for $\left( m,n\right) =\left( 1,1\right) $ and $z=0.6$ in our scheme,
including external and internal losses shown in Figs. 13(a) and 13(b),
respectively. For this season, in the following simulations, we shall give
more detailed analysis for our scheme about the effects of photon losses on
the superresolution and the phase sensitivity. For simplicity, the relevant
calculation details are not shown here, please refer to the appendix A.
\begin{figure}[tbp]
\label{Fig13} \centering \includegraphics[width=8cm]{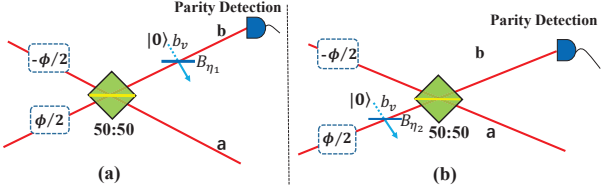}
\caption{{}(Color online) Schematic diagram of the photon losses (a) in
front of the parity detection (denoted as an external loss) and (b) between
the phase shifter and the second BS (denoted as an internal loss).}
\end{figure}
\begin{figure}[tbp]
\label{Fig14} \centering \includegraphics[width=0.8\columnwidth]{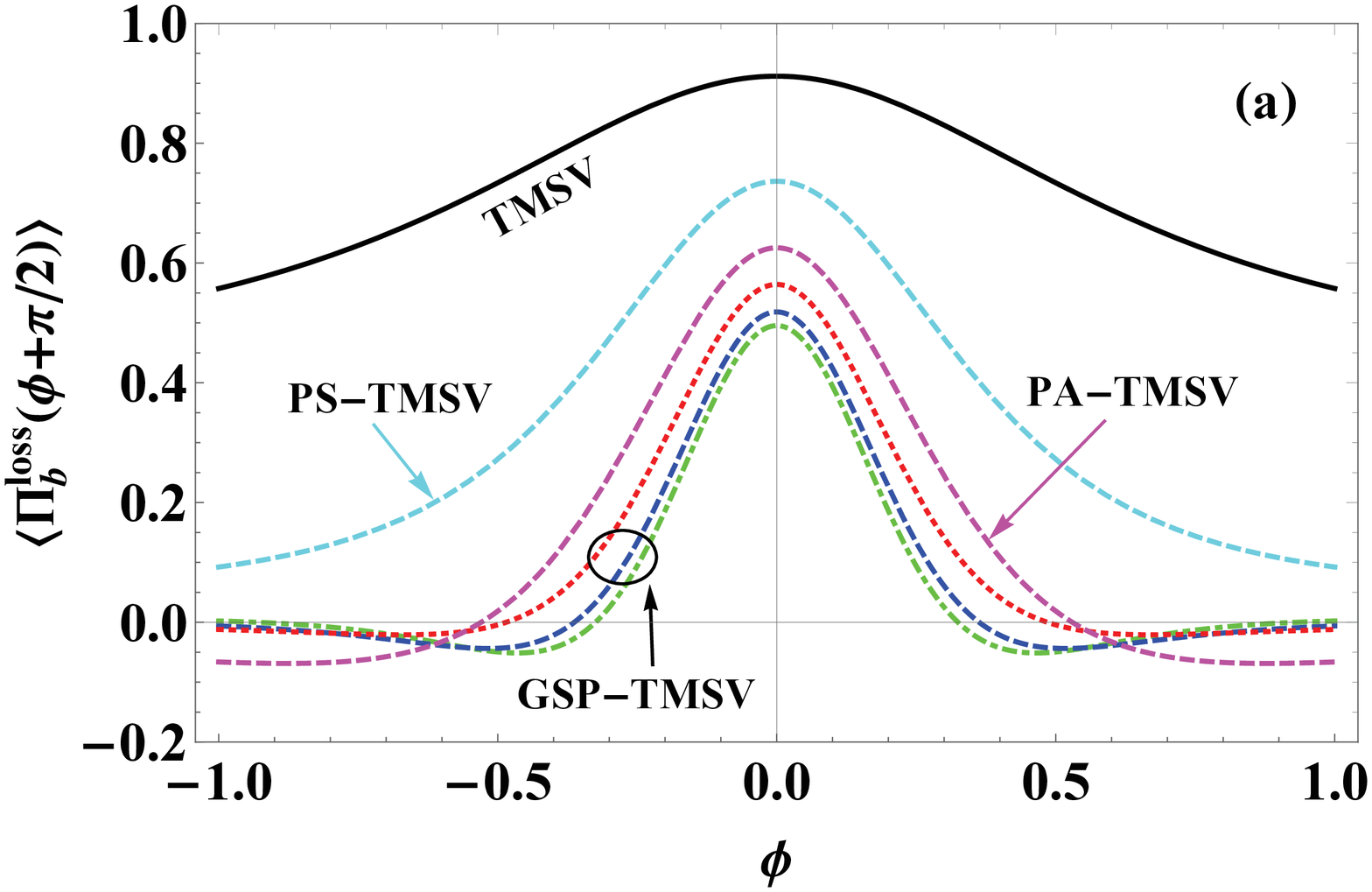} %
\includegraphics[width=0.8\columnwidth]{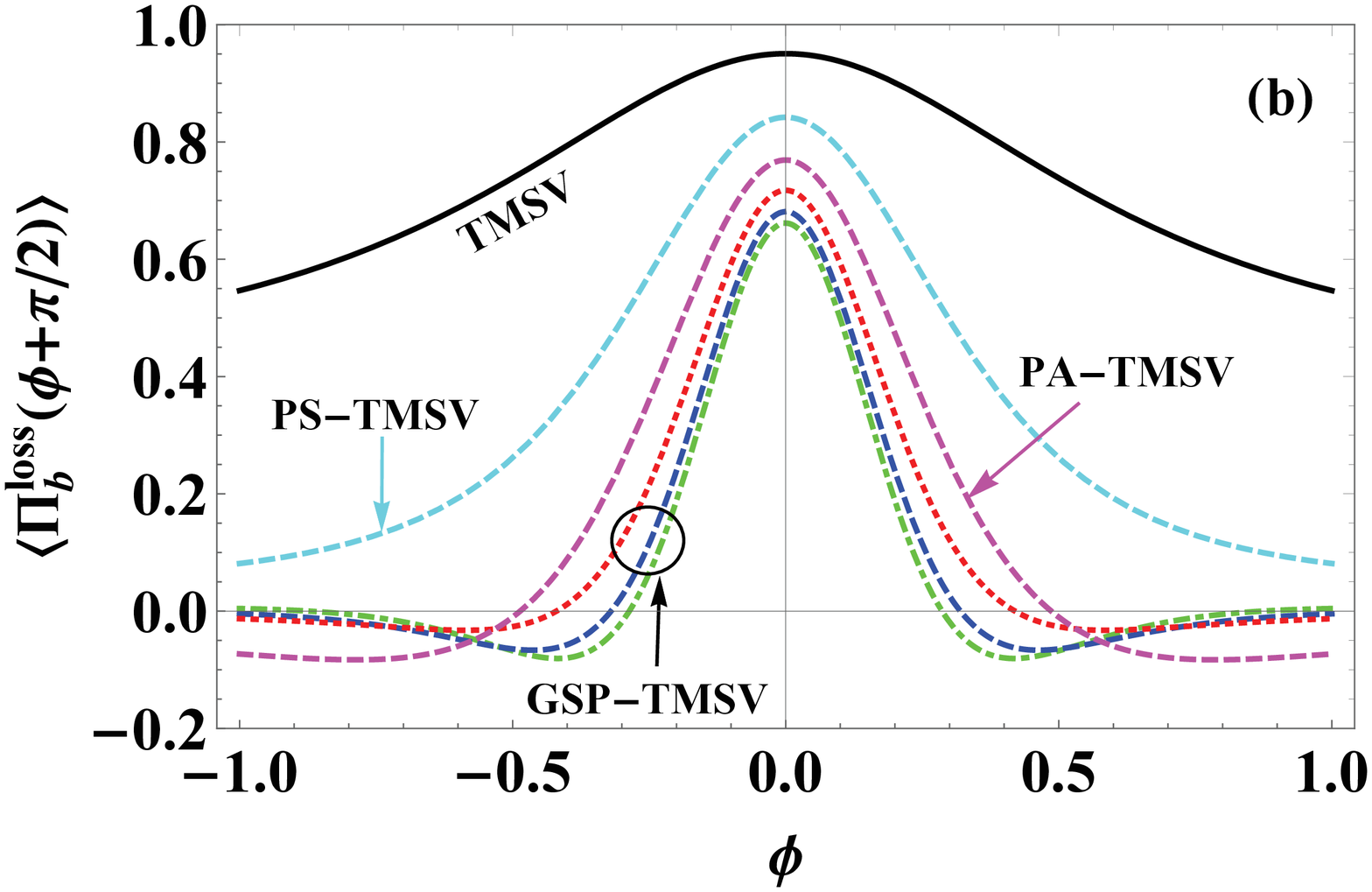}
\caption{{}(Color online) The expectation values of the parity operator $%
\left \langle \Pi _{b}(\protect \phi +\protect \pi /2)\right \rangle $ with
(a) external losses and (b) internal losses as a function of $\protect \phi $
for some fixed parameter $z=0.6$, $\protect \eta _{1}=\protect \eta _{2}=0.9$
and $m=n=1$. The dot-dashed lines represent the our work for several
different $s=0,0.5,1$ (corresponding to green, red, and blue dot-dashed
line, respectively). As a comparison, dashed lines represent the previous
work performing the PA (magenta color line) and the PS operations (cyan
color line). Solid line corresponds to the TMSV case.}
\end{figure}
\begin{figure}[tbp]
\label{Fig15} \centering \includegraphics[width=0.8\columnwidth]{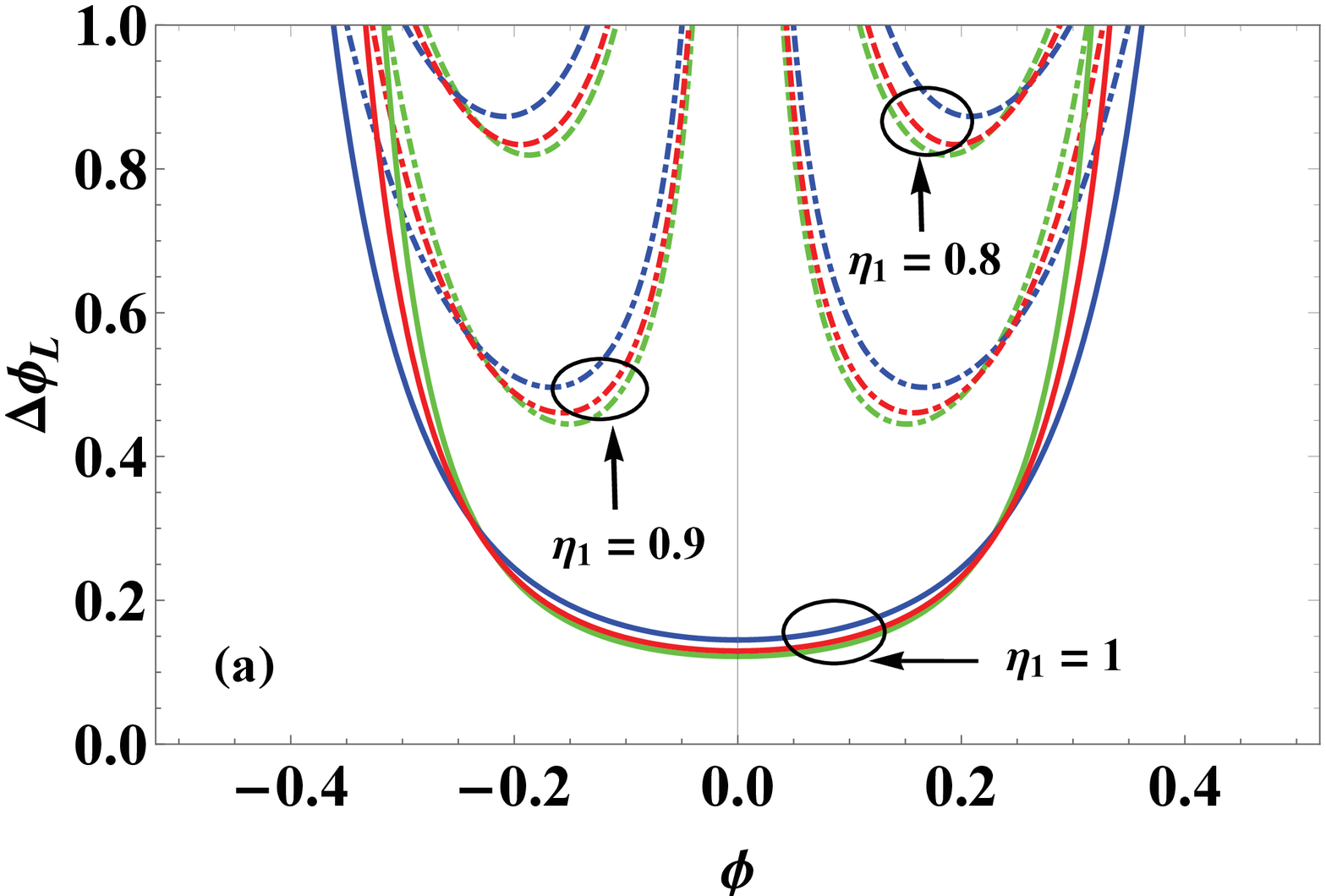} %
\includegraphics[width=0.8\columnwidth]{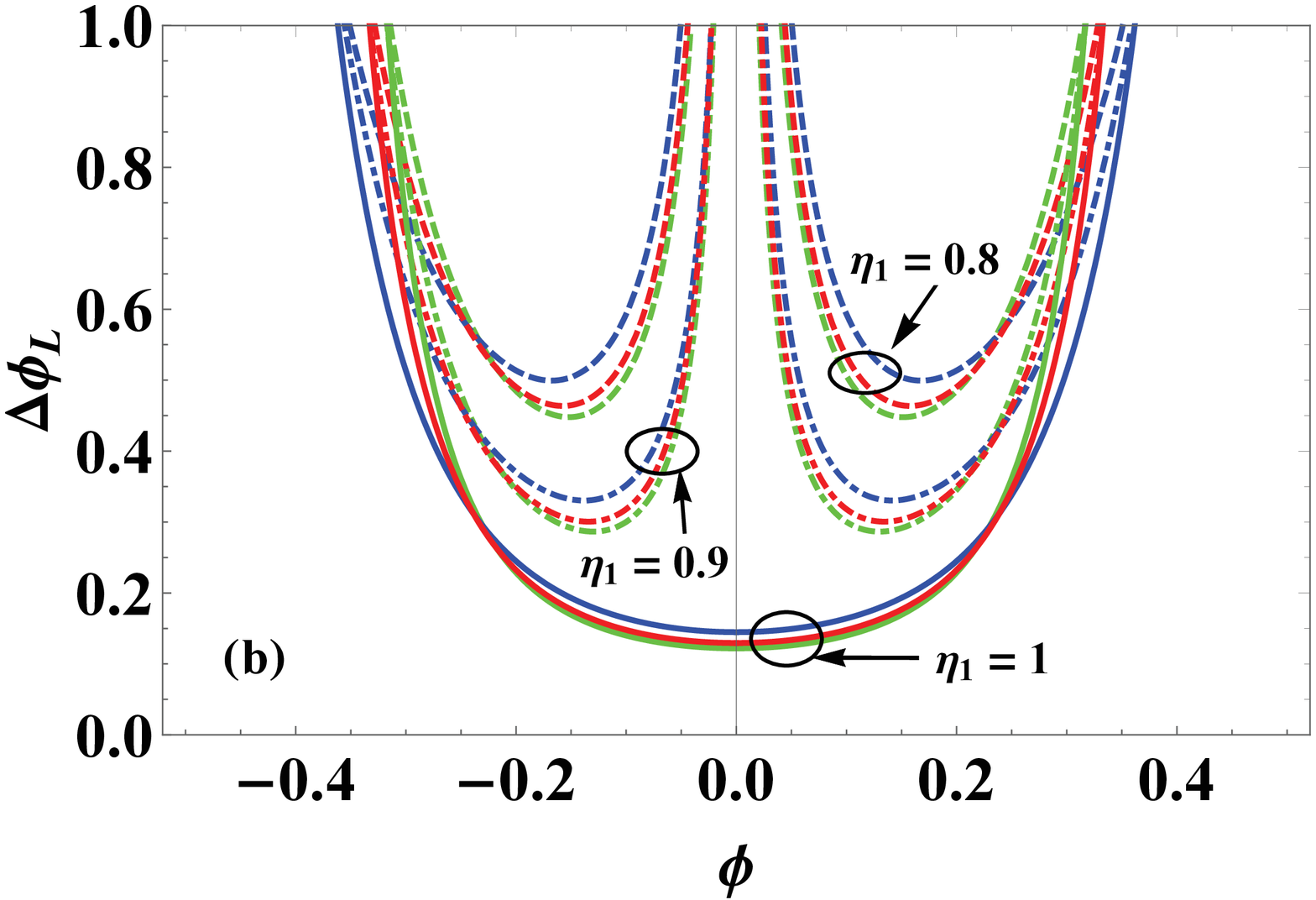}
\caption{{}(Color online) The phase sensitivity with (a) external losses and
(b) internal losses$\  \Delta \protect \phi _{L}$ as a function of $\protect%
\phi $ at some fixed parameter $z=0.6$ and $m=n=1$ for several dissipation
values $\protect \eta _{1}=\protect \eta _{2}=1,0.9,0.8$ and $s=0,0.5,1$
(corresponding to green, red, and blue dot-dashed lines, respectively.) As a
comparison, the solid line corresponds to the ideal cases, the dot-dashed
and dashed lines represent $\protect \eta _{1}=\protect \eta _{2}=0.9$ and $%
\protect \eta _{1}=\protect \eta _{2}=0.8$, respectively. }
\end{figure}

In Fig. 14, at a fixed dissipation value $\eta _{1}=\eta _{2}=0.9$, we show
the expectation values $\left \langle \Pi _{b}^{loss}(\phi +\pi
/2)\right
\rangle $ with the external- and internal- losses as a function
of the phase shift $\phi $ for several different parameters $s=0,0.5,1.$ It
is clear that the photon-loss processes make the central peak of $%
\left
\langle \Pi _{b}^{loss}(\phi +\pi /2)\right \rangle $ at $\phi =0$
lower than that of $\left \langle \Pi _{b}(\phi +\pi /2)\right \rangle $ for
the ideal cases (see Fig. 9). Nevertheless, we can see that the central
peaks of $\left
\langle \Pi _{b}^{loss}(\phi +\pi /2)\right \rangle $ at $%
\phi =0$ for all the GPS-TMSV inputs are much narrower than that for both
the TMSV and the single PA(PS)-TMSVs inputs, which reveals that the GPS
operations, especially for PS-then-PA $(s=0),$ help to increase the
superresolution even in the presence of photon losses, compared to both PA
and PS. Besides, in contrast to the external-loss cases, the central peaks
of $\left \langle \Pi _{b}^{loss}(\phi +\pi /2)\right \rangle $ for the
internal losses at $\phi =0 $ are relatively narrower, which implies that
the external losses have a greater influence on the superresolution than the
internal ones.

\begin{figure}[tbp]
\label{Fig16} \centering \includegraphics[width=0.8\columnwidth]{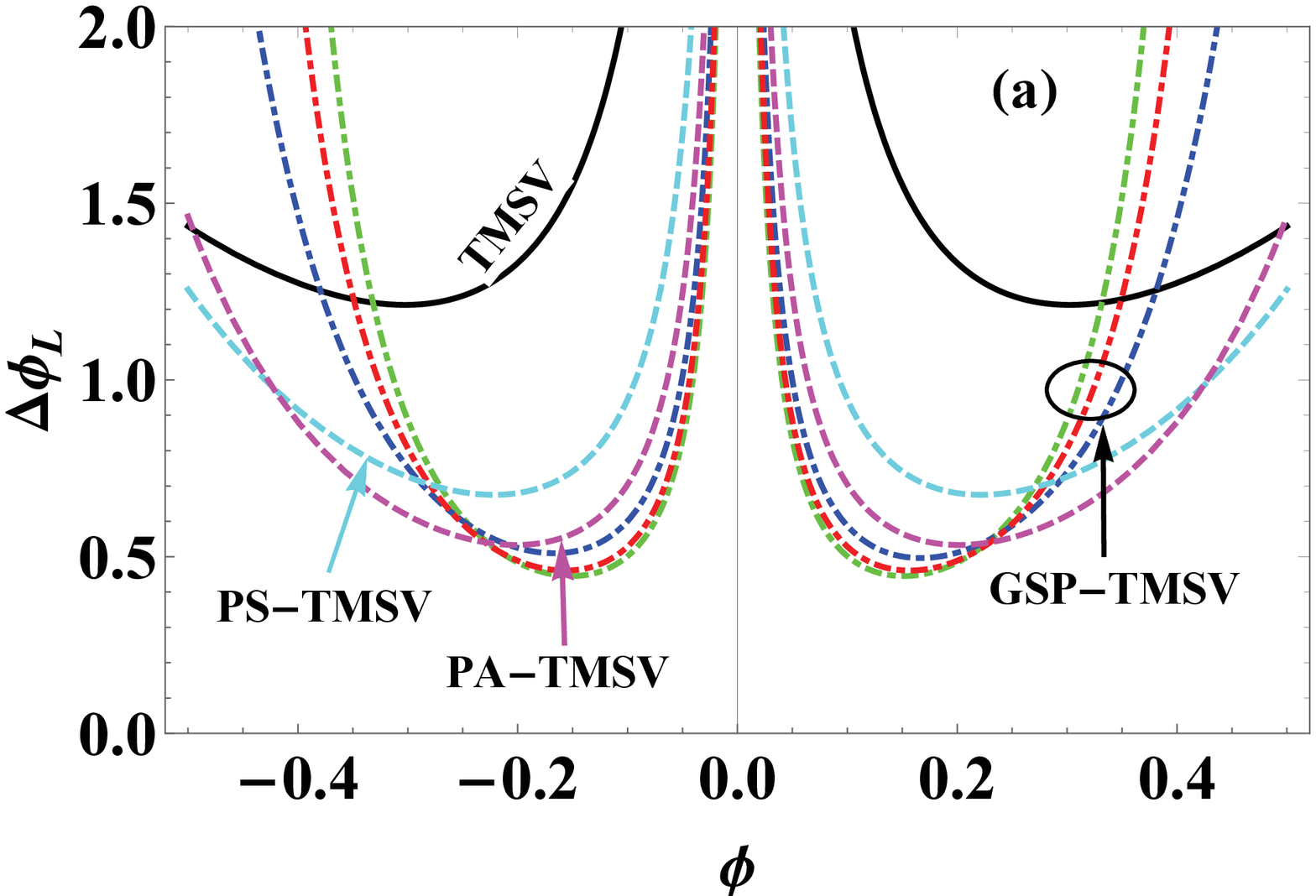} %
\includegraphics[width=0.8\columnwidth]{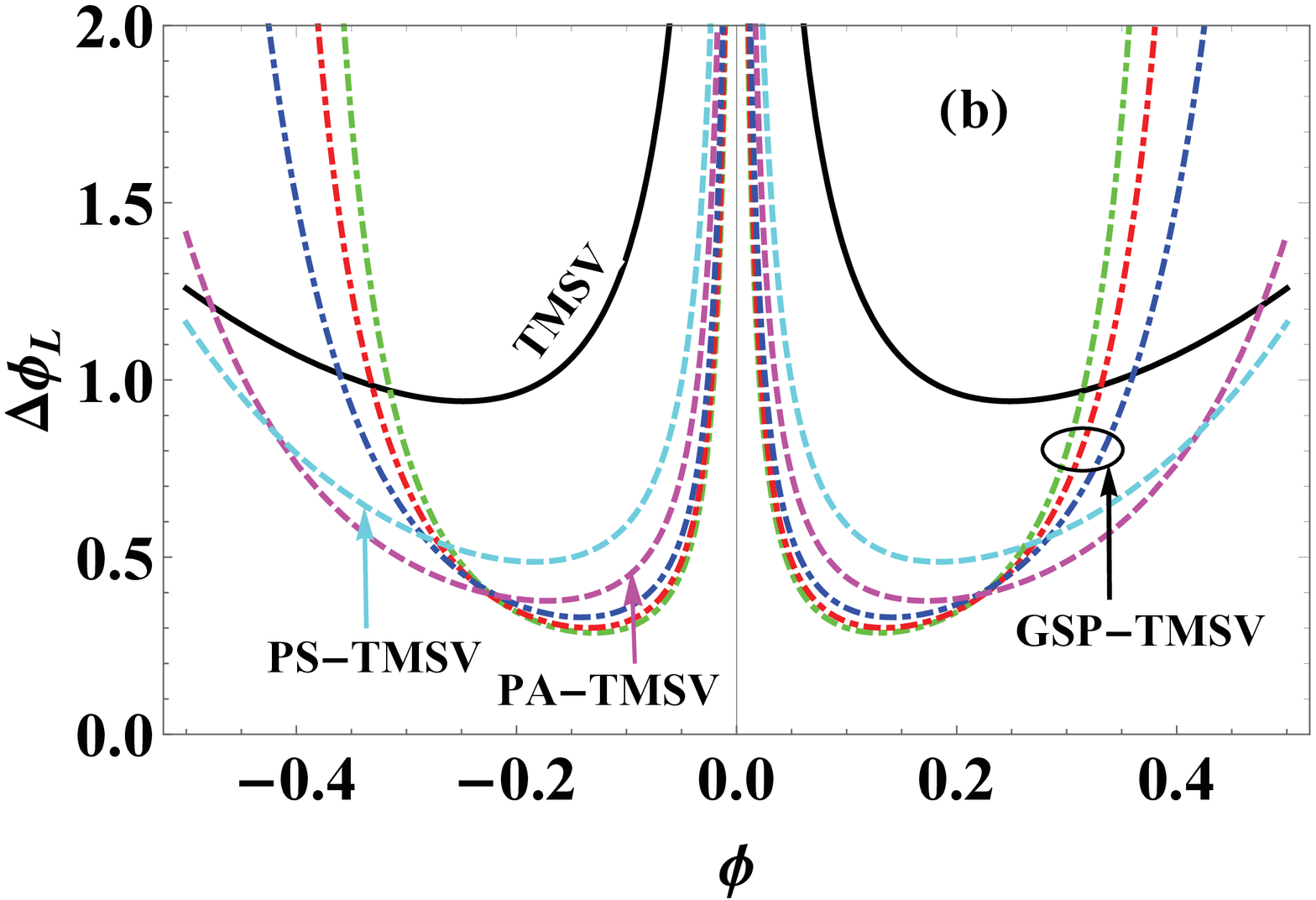}
\caption{{}(Color online) The phase sensitivity$\  \Delta \protect \phi _{L}$
with (a) external losses and (b) internal losses as a function of $\protect%
\phi $ for some fixed parameters $z=0.6,m=n=1$ and $\protect \eta _{1}=%
\protect \eta _{2}=0.9$. The dot-dashed lines represent the our work for
several different $s=0,0.5,$ and $1$ (corresponding to green, red, and blue
dot-dashed lines, respectively). As a comparison, dashed lines represent the
previous work performing the PA (magenta color line) and PS operations (cyan
color line). Black solid line corresponds to the TMSV case.}
\end{figure}

\begin{figure}[tbp]
\label{Fig17} \centering \includegraphics[width=0.8\columnwidth]{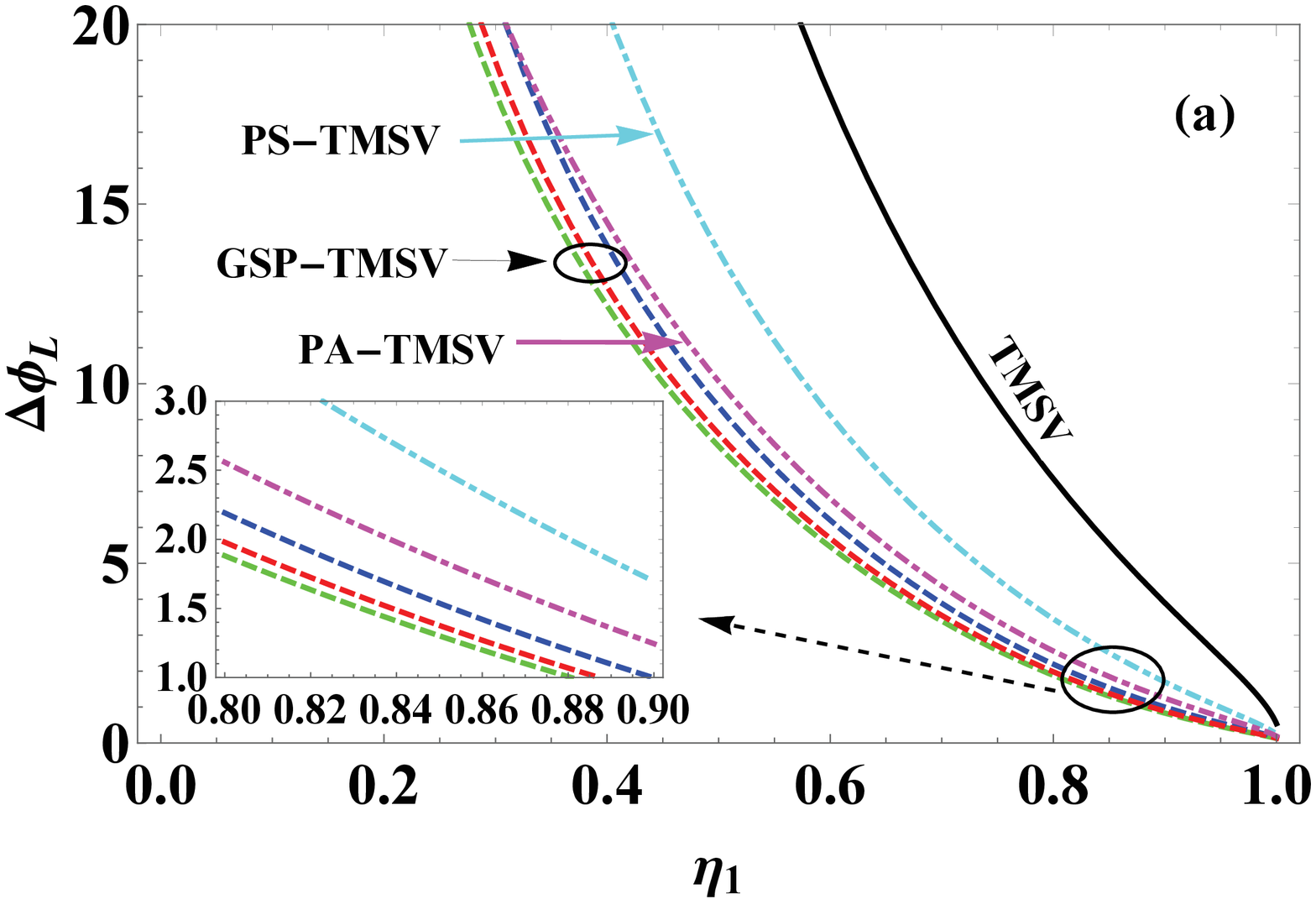} %
\includegraphics[width=0.8\columnwidth]{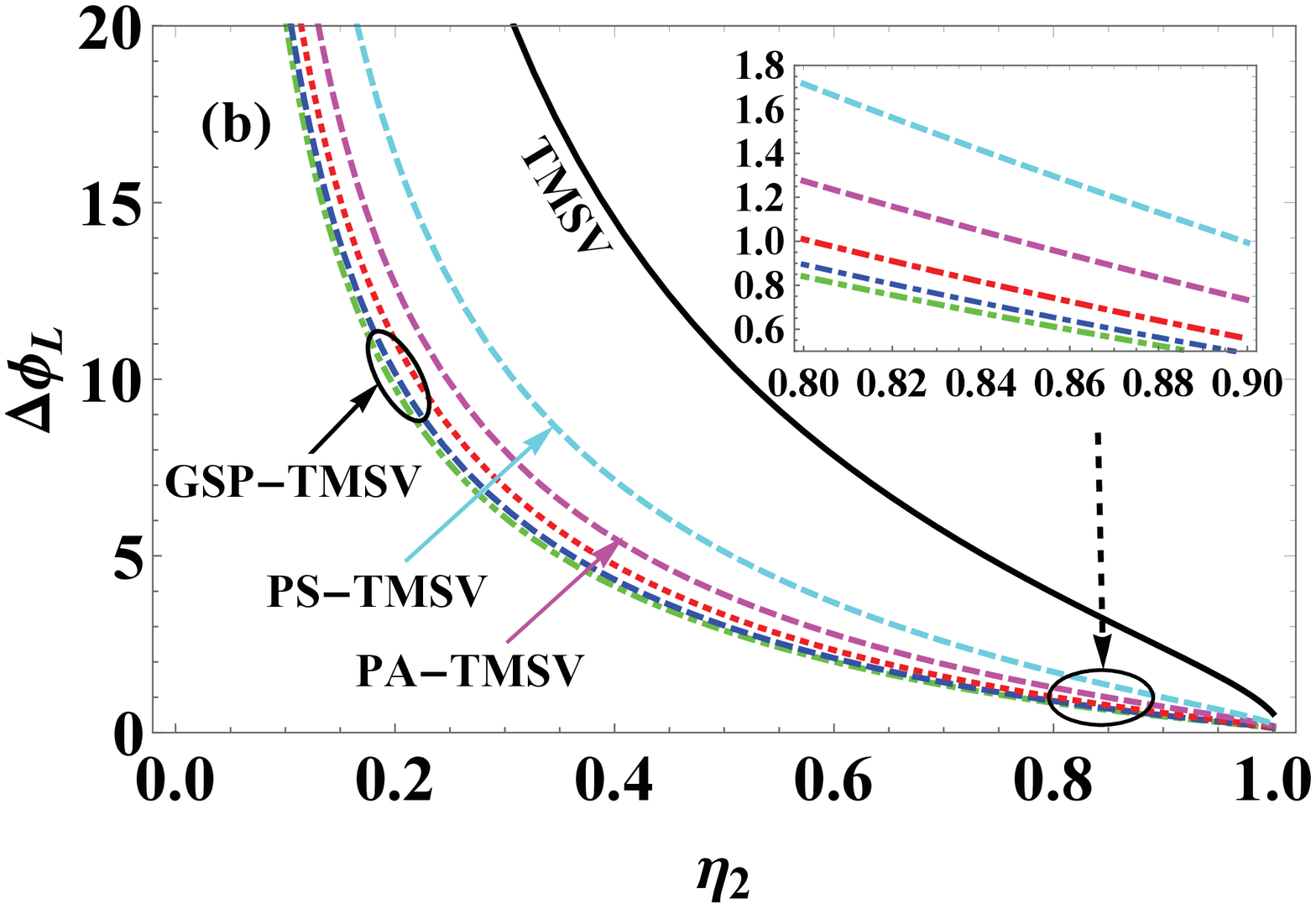}
\caption{{}(Color online) The phase sensitivity with (a) external-losses and
(b) internal-losses as a function of transmissivity of fictitious beam
splitter $\protect \eta _{1}($or $\protect \eta _{2})$ for some fixed
parameters $z=0.6,$ $\protect \phi =0.05$ and $s=0,0.5,1$ (green, red, and
blue dot-dashed lines, respectively). As a comparison, black solid line
corresponds to the TMSV case. Cyan- and purple dashed lines correspond to
the PS-TMSV and the PA-TMSV, respectively. }
\end{figure}

To visually display the effects of photon losses on phase sensitivity, we
illustrate the phase sensitivity$\  \Delta \phi _{L}$ as a function of the
phase $\phi $ for several dissipation values $\eta _{l}=1,0.9,0.8$ $(l=1,2),$
as shown in Fig. 15. The solid lines represent the ideal case with $\eta
_{l}=1$ where the optimal phase point is at $\phi _{opt}=0$. However, in the
presence of photon losses, the optimal phase point that tends to be far away
from zero for $\eta _{l}=$ $0.9,0.8$ is at $\phi _{opt}\neq 0,$ which leads
to the decrease of phase sensitivity. The reason may be that the noise could
be suppressed in near decorrelation point ($\phi =0$), as shown in Ref. \cite%
{57}. Furthermore, under the same accessible parameters except for $\eta
_{l}=1$, the phase sensitivity $\Delta \phi _{L}$ for the internal losses
performs better than that for the external-loss cases, which indicates that
the latter has a greater impact on the precision of phase measurement. In
order to show the advantages of our scheme, on the other hand, we take a
fixed $\eta _{l}=0.9$ and make a comparison about $\Delta \phi _{L}$
changing with the phase $\phi $ for several non-Gaussian resources inputs
involving single PA(PS)-TMSVs and GSP-TMSV, as shown in Fig. 16. It is found
that, compared with the TMSV input (black solid line), these non-Gaussian
resources can still be used for enhancing the phases sensitivity even in the
presence of photon losses. Among them, all the GSP-TMSV inputs, present
better advantages for further improving the phases sensitivity when
considering photon losses, in which the PS-then-PA $(s=0)$ is the best.

\begin{figure}[tbp]
\label{Fig18} \centering \includegraphics[width=0.8\columnwidth]{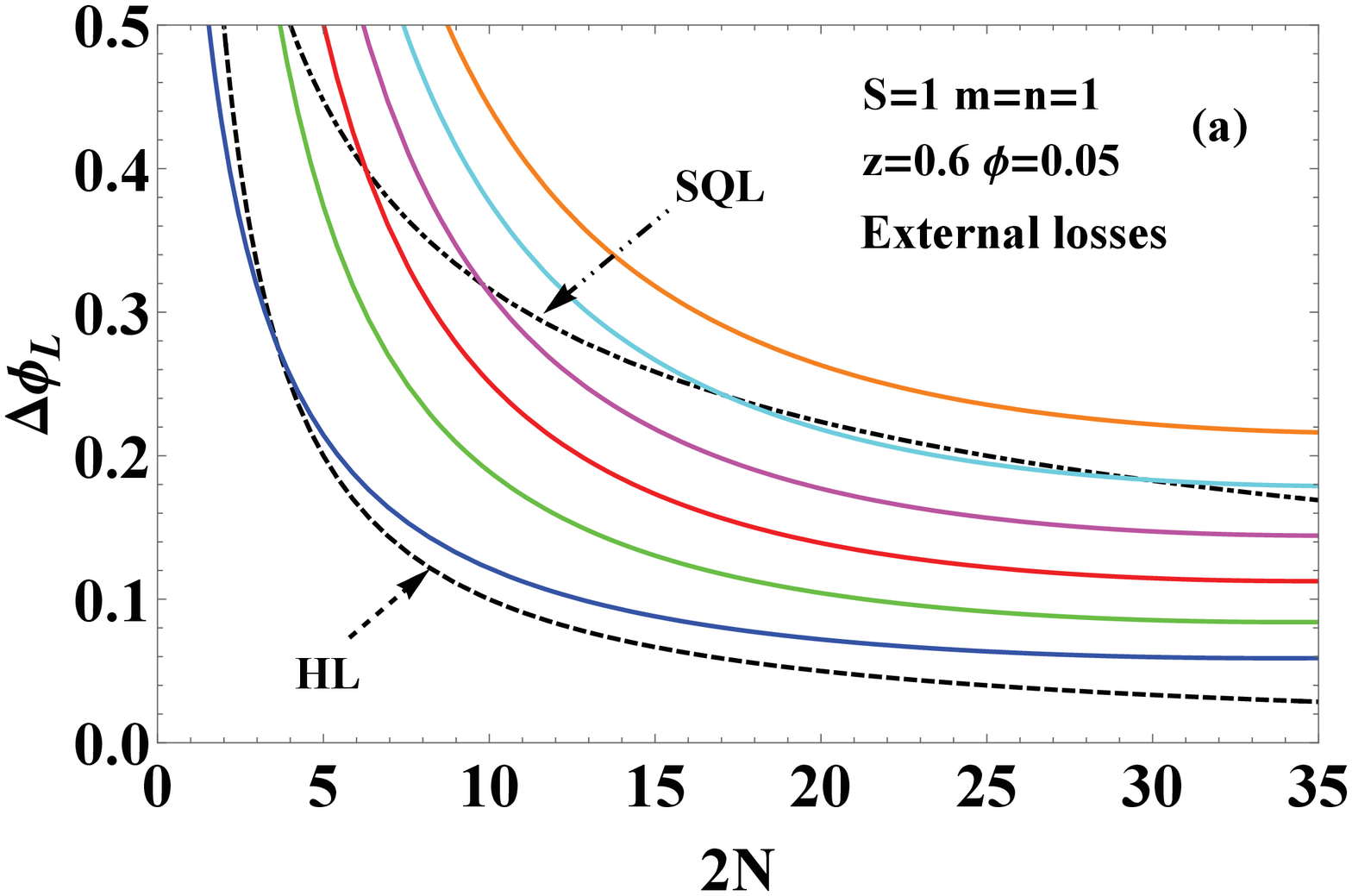} %
\includegraphics[width=0.8\columnwidth]{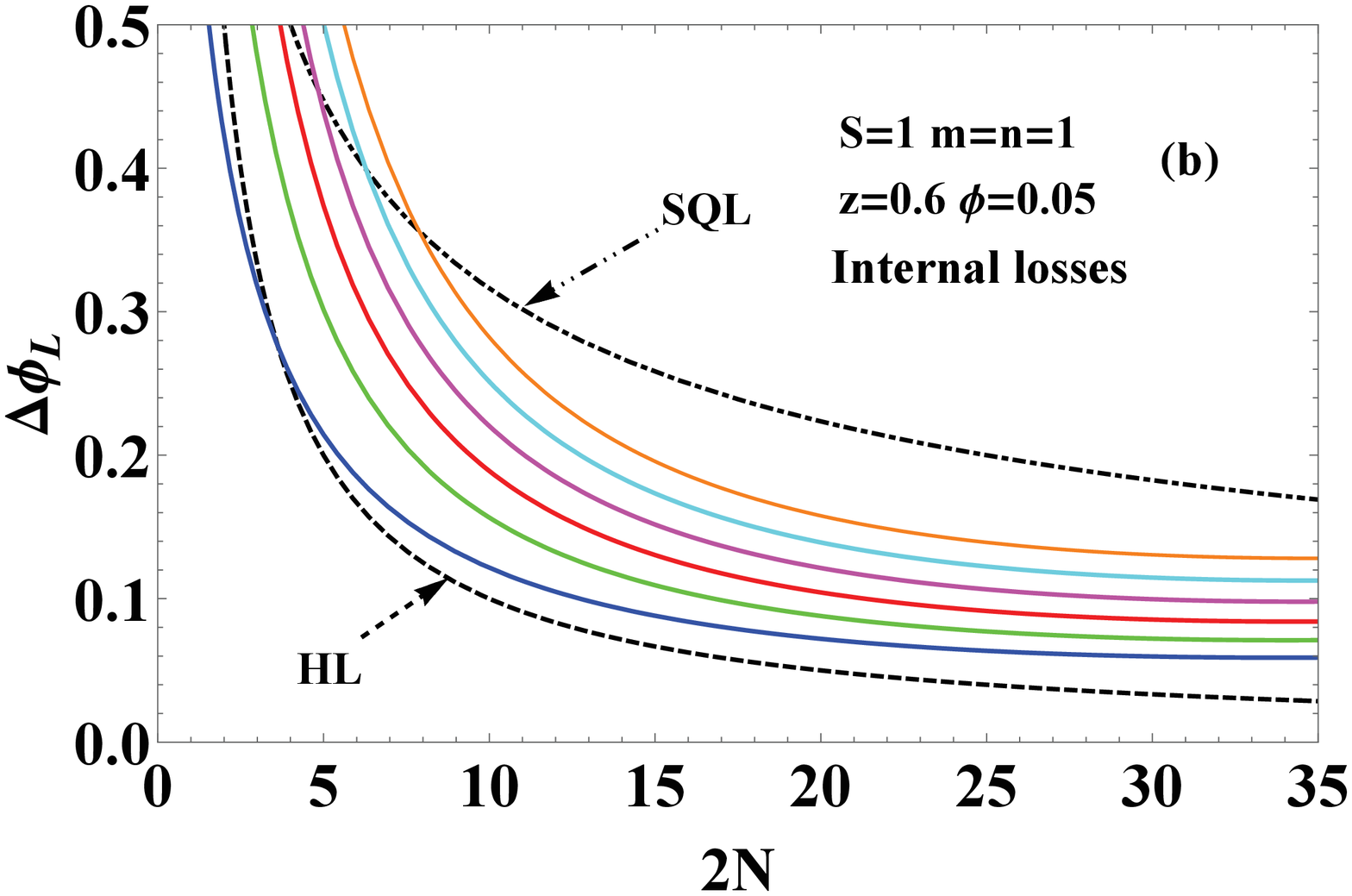}
\caption{{}(Color online) The phase sensitivity$\  \Delta \protect \phi _{L}$
with (a) external-losses and (b) internal-losses as a function of the total
APN $2N$ at some fixed parameters $s=1,m=n=1$ and $\protect \phi =0.05$. The
dot-dashed and dashed line correspond to the SQL and HL, respectively. The
color lines from down to up correspond to dissipation value $\protect \eta %
_{1}=\protect \eta _{2}=1,0.99,0.98,0.97,0.96,$ and $0.95$, respectively.}
\end{figure}

In Fig. 17, we plot the phase sensitivity $\Delta \phi _{L}$ as a function
of $\eta _{1}($or $\eta _{2})$ for several non-Gaussian resources inputs
mentioned above at fixed parameters $z=0.6$ and $\phi =0.05$, from which the
phase sensitivity can be deteriorated severely with the decrease of $\eta
_{1}($or $\eta _{2}).$ In contrast to the TMSV input, fortunately, the
phases sensitivity can be still improved even in the presence of photon
losses by using these non-Gaussian resources, especially for the GSP-TMSV.
In this sense, this means that the GSP operations are more effective to
resist photon losses comparing with the PA(PS) operation. In addition, the
effects of the external losses on phase sensitivity are more serious than
the internal-loss cases, particularly in the small $\eta _{1}($or $\eta
_{2}) $ regimes.

On the other hand, as shown in Fig. 12, without losses, it is shown that the
SQL can be broken for all the GSP-TMSV inputs and the HL for the cases of $%
s=0.5,1$ can be beaten in the regime of the small total APN. In the context
of photon losses, then, can the two limits be broken by using the GSP-TMSV?
To this end, for some given parameters $\left( m,n\right) =\left( 1,1\right)
,z=0.6$ and $\phi =0.05,$ in Fig. 18, we plot the phase sensitivity $\Delta
\phi _{L}$ as a function of total APN $2N$ for several dissipation values $%
\eta _{l}=1,0.99,0.98,0.97,0.96$ and $0.95$. It is clearly seen that the
phase sensitivity decreases rapidly with the decrease of $\eta _{1}($or $%
\eta _{2}).$ Particularly, when $\eta _{1}=0.95,$ the SQL cannot be achieved
for the external-loss cases but can be still broken through at large range
of the total APN for the internal-loss ones. These results indicate that the
external losses make against to the effective improvement of phase
sensitivity compared to the internal-loss cases.

\section{Conclusion}

In summary, we propose a scheme to improve the phase sensitivity and
resolution using a novel non-Gaussian quantum state, the GSP-TMSV, as the
input of the MZI via parity detection. The nonclassicality of the
proposed state is discussed in terms of the APN, the anti-bunching
effect and two-mode squeezing property. We also investigate both the QFI and the phase
resolution/sensitivity based on parity detection when using GSP-TMSV as input  in
detail. The numerical results show that our scheme, especially for the case
of the PS-then-PA TMSV, is always superior to the original TMSV scheme in
terms of the QFI and the phase resolution and sensitivity, which is caused
by the fact that the total APN of the former is larger than that of the
latter.

In addition, to show the advantages of our scheme, we also make
comparisons between the GSP-TMSV and the previous PA(or PS)-TMSV schemes
in terms of the total APN, the QFI, and the phase resolution and sensitivity. The
results indicate that the current scheme can surpass the previous schemes, especially when the PS-then-PA TMSV is used. This means that
the proposed GSP operation can obviously improve the QFI and the phase
resolution and sensitivity. In addition, comparing with the
single-side GSP operations, the improvement of phase sensitivity via two-side symmetric ones is
more remarkable under the same accessible parameters. Furthermore, the SQL can always be surpassed by our scheme and the HL can be
beaten when $s=0.5,1$ in the regime of the small total APN, but
not by the case of $s=0$. These results show that the GSP-TMSV is an useful
resource for improving phase sensitivity remarkably beyond the classical
limit, and even going beyond the HL.

From a realistic point of view, we also study the sensitivity of phase
estimation with parity detection in the presence of photon losses, including
external- and internal- losses. The results indicate that compared with the
internal photon losses, the external ones have a greater impact on phase
sensitivity when several non-Gaussian resources, involving single
PA(PS)-TMSVs and GSP-TMSV, are used as the inputs. Dramatically, under the
same parameters, the phase sensitivity with  the GSP-TMSV, especially for
the case of $s=0$, can be better than those base on
the TMSV or the PA(PS)-TMSV in the presence of photon losses. Besides, it is also noted that in the presence of photon losses,
the HL cannot be beaten, but fortunately the SQL can still be surpassed when GSP-TMSV is used as the inputs
particularly when  the total APN is large. Our results shown here can find important applications in quantum metrology.

\begin{acknowledgments}
This work is supported by the National Natural Science Foundation of China
(Grant Nos. 11964013,11664017), the Training Program for Academic and
Technical Leaders of Major Disciplines in Jiangxi Province, and Z.L. is
supported by a startup Grant (No.74130-18841222) at Sun Yat-sen University.
\end{acknowledgments}

\textbf{Appendix\ A: Derivation of the phase sensitivity with parity
detection in the presence of photon losses}

In order to derive the phase sensitivity with parity detection in the
presence of photon losses, for simplicity, here we consider two special
photon-loss processes, i.e., the external loss and the internal one, shown
in Fig. 13. In practice, the photon losses on auxiliary mode $b_{v}$ can be
structured using a fictitious beam splitter (denoted as $B_{\eta _{i}}$)
with a dissipation factor $\eta _{i}$ ($i=1$ and $2$ corresponding to the
external and internal losses, respectively), whose transform relation is
given by \cite{58a}%
\begin{equation}
B_{\eta _{i}}^{\dagger }\binom{b}{b_{v}}B_{\eta _{i}}=\left(
\begin{array}{cc}
\sqrt{\eta _{i}} & \sqrt{1-\eta _{i}} \\
-\sqrt{1-\eta _{i}} & \sqrt{\eta _{i}}%
\end{array}%
\right) \binom{b}{b_{v}}.  \tag{A1}
\end{equation}%
It is worth mentioning that the smaller the values of $\eta _{i}$, the more
severe the photon losses. Particularly, $\eta _{i}=1$ corresponds to the
ideal case. To get the parity operator in the presence of the external
losses, on one hand, it is necessary to rewrite Eq. (\ref{20}) under the
Weyl ordering representation \cite{58}, i.e.,%
\begin{equation}
\Pi _{b}=\frac{\pi }{2}%
\begin{array}{c}
: \\
:%
\end{array}%
\delta \left( b\right) \delta \left( b^{\dagger }\right)
\begin{array}{c}
: \\
:%
\end{array}%
,  \tag{A2}
\end{equation}%
where $%
\begin{array}{c}
: \\
:%
\end{array}%
\bullet
\begin{array}{c}
: \\
:%
\end{array}%
$ denotes the symbol of the Weyl ordering and $\delta \left( \bullet \right)
$ denotes the delta function. Thus, by using Eq. (A1), one can obtain the
parity operator with external losses (denoted as $\Pi _{b}^{loss}$), namely,
\begin{subequations}
\begin{align}
\Pi _{b}^{loss}& \text{=}\frac{\pi }{2}_{v}\left \langle 0\right \vert
\begin{array}{c}
: \\
:%
\end{array}%
\delta \left( \sqrt{\eta _{i}}b+\sqrt{1-\eta _{i}}b_{v}\right)   \notag \\
& \times \delta \left( \sqrt{\eta _{i}}b^{\dagger }+\sqrt{1-\eta _{i}}%
b_{v}^{\dagger }\right)
\begin{array}{c}
: \\
:%
\end{array}%
\left \vert 0\right \rangle _{v},  \tag{A3}
\end{align}%
where $\left \vert 0\right \rangle _{v}$ is the vacuum noise input on
auxiliary mode $b_{v}$. Finally, according to the classical correspondence
of the operator
\end{subequations}
\begin{subequations}
\begin{align}
\begin{array}{c}
: \\
:%
\end{array}%
f\left( b,b^{\dagger },b_{v},b_{v}^{\dagger }\right)
\begin{array}{c}
: \\
:%
\end{array}%
& \text{=}4\int d^{2}\beta d^{2}\gamma f\left( \beta ,\beta ^{\ast },\gamma
,\gamma ^{\ast }\right)   \notag \\
& \times \Delta \left( \beta ,\beta ^{\ast }\right) \Delta \left( \gamma
,\gamma ^{\ast }\right) ,  \tag{A4}
\end{align}%
with Wigner operators under the normal ordering \cite{59}
\end{subequations}
\begin{subequations}
\begin{align}
\Delta \left( \beta ,\beta ^{\ast }\right) & =\colon \exp \left[
-2(b^{\dagger }-\beta ^{\ast })(b-\beta )\right] \colon ,  \notag \\
\Delta \left( \gamma ,\gamma ^{\ast }\right) & =\colon \exp \left[
-2(b_{v}^{\dagger }-\gamma ^{\ast })(b_{v}-\gamma )\right] \colon ,  \tag{A5}
\end{align}%
and using the IWOP technique \cite{60}, it is easy to obtain
\end{subequations}
\begin{equation}
\Pi _{b}^{loss}=\colon e^{-2\eta _{1}b^{\dagger }b}\colon =\left( 1-2\eta
_{1}\right) ^{b^{\dagger }b},  \tag{A6}
\end{equation}%
where the symbol $\colon \colon $ denotes the normal ordering. Thus,
combining Eqs. (\ref{14}) and (A6), the average value of $\Pi _{b}^{loss}$
for the output state can be given by
\begin{equation}
\left \langle \Pi _{b}^{loss}\right \rangle =\text{Tr}\left[ \rho _{out}\Pi
_{b}^{loss}\right] =\widetilde{\Re }\frac{uu_{1}\sqrt{\vartheta _{1}}}{P_{d}%
\sqrt{\vartheta _{2}^{2}-\vartheta _{3}}},  \tag{A7}
\end{equation}%
with
\begin{subequations}
\begin{align}
\vartheta _{1}& =1-vv_{1}\sin ^{2}\varphi ,  \notag \\
\vartheta _{2}& =1-vv_{1}+2\eta _{1}v_{1}v\cos ^{2}\varphi ,  \notag \\
\vartheta _{3}& =\left( 1-2\eta _{1}\right) ^{2}\sin ^{2}\varphi
v_{1}v\left( 1-vv_{1}\right) ^{2}.  \tag{A8}
\end{align}%
On the other hand, different from the derivation of Eq. (A6), we rewrite the
parity operator with the internal losses as
\end{subequations}
\begin{subequations}
\begin{align}
\widetilde{\Pi }_{b}^{loss}& \text{=}_{v}\left \langle 0\right \vert
B_{1}^{\dagger }U^{\dagger }\left( \varphi \right) B_{v}^{\dagger
}B_{2}^{\dagger }e^{i\pi b^{\dagger }b}B_{2}B_{v}U\left( \varphi \right)
B_{1}\left \vert 0\right \rangle _{v}  \notag \\
& \text{=}\colon e^{\varpi _{1}a^{\dagger }a-\varpi _{2}b^{\dagger }a-\varpi
_{2}^{\ast }a^{\dagger }b+\varpi _{3}b^{\dagger }b}\colon ,  \tag{A9}
\end{align}%
where $U\left( \varphi \right) $ is given in Eq. (\ref{12}) and
\end{subequations}
\begin{subequations}
\begin{align}
\varpi _{1}& =\sqrt{\eta _{2}}\cos \varphi -\frac{1+\eta _{2}}{2},  \notag \\
\varpi _{2}& =\frac{\left( \eta _{2}+1\right) ^{2}-4\eta _{2}\cos
^{2}\varphi }{4\left( i\eta _{2}-i+2\sqrt{\eta _{2}}\sin \varphi \right) },
\notag \\
\varpi _{3}& =-\sqrt{\eta _{2}}\cos \varphi -\frac{1+\eta _{2}}{2},
\tag{A10}
\end{align}%
as well as we have used the following transformation relationships
\end{subequations}
\begin{subequations}
\begin{align}
B_{1}^{\dagger }\left(
\begin{array}{c}
a \\
b%
\end{array}%
\right) B_{1}& =\frac{\sqrt{2}}{2}\left(
\begin{array}{cc}
1 & i \\
i & 1%
\end{array}%
\right) \left(
\begin{array}{c}
a \\
b%
\end{array}%
\right) ,  \notag \\
B_{2}^{\dagger }\left(
\begin{array}{c}
a \\
b%
\end{array}%
\right) B_{2}& =\frac{\sqrt{2}}{2}\left(
\begin{array}{cc}
1 & -i \\
-i & 1%
\end{array}%
\right) \left(
\begin{array}{c}
a \\
b%
\end{array}%
\right) .  \tag{A11}
\end{align}%
Thus, for a given input GSP-TMSV, one can obtain the expectation value of $%
\widetilde{\Pi }_{b}^{loss}$ for the internal losses, i.e.,
\end{subequations}
\begin{subequations}
\begin{align}
\left \langle \Pi _{b}^{loss}\right \rangle & =\mathtt{Tr}[\left \vert \psi
\right \rangle _{ab}\left \langle \psi \right \vert \widetilde{\Pi }_{b}^{loss}]
\notag \\
& =\frac{\widetilde{\Re }uu_{1}}{p_{d}}\left[ \left( 1-\omega _{1}\right)
^{2}-\omega _{2}\right] ^{-\frac{1}{2}},  \tag{A12}
\end{align}%
with
\end{subequations}
\begin{subequations}
\begin{align}
\omega _{1}& =v_{1}v\left( \varpi _{1}\varpi _{3}-\eta _{2}+\left \vert
\varpi _{2}\right \vert ^{2}\right) ,  \notag \\
\omega _{2}& =4\left \vert \varpi _{2}\right \vert ^{2}v_{1}^{2}v^{2}\left(
\varpi _{1}\varpi _{3}-\eta _{2}\right) .  \tag{A13}
\end{align}%
Finally, using Eqs. (A7) and (A12), the phase sensitivity (denoted as $%
\Delta \phi _{L}$) in the presence of external and internal losses can be
estimated by the error propagation formula

\end{subequations}
\begin{equation}
\Delta \phi _{L}=\frac{\sqrt{1-\left \langle \Pi _{b}^{loss}\right \rangle
^{2}}}{\left \vert \partial \Pi _{b}^{loss}/\partial \phi \right \vert }.
\tag{A14}
\end{equation}


\begin{thebibliography}{99}
\bibitem{1} S. L. Braunstein and C. M. Caves, Statistical Distance and the
Geometry of Quantum States, Phys. Rev. Lett. 72, 3439 (1994).

\bibitem{2} J. P. Dowling, Quantum optical metrology-the lowdown on
high-N00N states, Contem. Phys. 49, 125 (2008).

\bibitem{3} J. J . Bollinger, W. M. Itano, and D. J. Wineland, Optimal
frequency measurements with maximally correlated states, Phys. Rev. A 54,
4649 (1996)

\bibitem{4} V. Giovannetti, S. Lloyd, and L. Maccone, Advances in quantum
metrology, Nat. Photonics 5, 222 (2011).

\bibitem{5} C. M. Caves, Quantum-mechanical noise in an interferometer,
Phys. Rev. D 23, 1693 (1981).

\bibitem{6} R. Carranza and C. C. Gerry, Photon-subtracted two-mode squeezed
vacuum states and applications to quantum optical interferometry, J. Opt.
Soc. Am. B 29, 2581 (2012).

\bibitem{7} H. Kwon, K. C. Tan, T. Volkoff, and H. Jeong, Nonclassicality as
a Quantifiable Resource for Quantum Metrology, Phys. Rev. Lett. 122, 040503
(2019).

\bibitem{8} P. M. Anisimov, G. M. Raterman, A. Chiruvelli, W. N. Plick, S.
D. Huver, Quantum Metrology with Two-Mode Squeezed Vacuum: Parity Detection
Beats the Heisenberg Limit, Phys. Rev. Lett. 104, 103602 (2010).

\bibitem{9} I. Afek, O. Ambar, and Y. Silberberg, High-NOON States by Mixing
Quantum and Classical Light, Science 328, 879 (2010).

\bibitem{10} V. Giovannetti, S. Lloyd, and L. Maccone, Quantum-Enhanced
Measurements: Beating the Standard Quantum Limit, Science 306, 1330 (2004).

\bibitem{11} J. Joo, W. J. Munro, and T. P. Spiller, Quantum Metrology with
Entangled Coherent States, Phys. Rev. Lett. 107, 083601 (2011).

\bibitem{12} M. Jarzyna and R. D. Dobrza\'{n}ski, Quantum interferometry
with and without an external phase reference Phys. Rev. A 85, 011801 (2012).

\bibitem{13} J. J. Cooper, D. W. Hallwood, J. A. Dunningham, and J. Brand,
Robust Quantum Enhanced Phase Estimation in a Multimode Interferometer,
Phys. Rev. Lett. 108, 130402 (2012).

\bibitem{14} T. Eberle, V. Hadchen, and R. Schnabel, Stable control of 10 dB
two-mode squeezed vacuum states of light, Opt. Express 21, 11546 (2013).

\bibitem{15} N. Namekata, Y. Takahashi, G. Fujii, D. Fukuda, S. Kurimura,
and S. Inoue, Non-Gaussian operation based on photon subtraction using a
photon-number-resolving detector at a telecommunications wavelength, Nature
Photonics 10, 1038 (2010).

\bibitem{16} L. Y. Hu, Z. Y. Liao, and M. S. Zubairy, Continuous-variable
entanglement via multiphoton catalysis, Phys. Rev. A 95, 012310 (2017).

\bibitem{17} L. Y. Hu, M. Al-amri, Z. Y. Liao, and M. S. Zubairy,
Entanglement improvement via a quantum scissor in a realistic environment,
Phys. Rev. A 100, 052322 (2019).

\bibitem{18} G. S. Agarwal and K. Tara, Nonclassical properties of states
generated by the excitations on a coherent state. Phys. Rev. A 43, 492
(1991).

\bibitem{19} A Zavatta, V. Parigi, and M Bellini, Experimental
nonclssicality of single-photon-added thermal light states. Phys. Rev. A 75,
052106 (2007).

\bibitem{20} L. Y. Hu and Z. M. Zhang, Statistical properties of coherent
photon-added two-mode squeezed vacuum and its inseparability, J. Opt. Soc.
Am. B 30, 518 (2013)

\bibitem{21} A. Zavatta, S. Viciani, and M. Bellini, Quantum-to-classical
transition with single-photon-added coherent states of light,\ Science
\textbf{306}, 660--662 (2004).

\bibitem{22} Y. Ouyang, S. Wang, and L. J. Zhang, Quantum optical
interferometry via the photon added two-mode squeezed vacuum states, J. Opt.
Soc. Am. B 33, 1373 (2016).

\bibitem{23} S. Y. Lee, S. W. Ji, H. J. Kim, and H. Nha, Enhancing quantum
entanglement for continuous variables by a coherent superposition of photon
subtraction and addition, Phys. Rev. A\ 84, 012302 (2011).

\bibitem{24} S. Wang, L. L. Hou, X. F. Chen, and X. F. Xu,
Continuous-variable quantum teleportation with non-Gaussian entangled states
generated via multiple-photon subtraction and addition, Phys. Rev. A 91,
063832 (2015).

\bibitem{25} E. D. Lopaeva, I. R. Berchera, I. P. Degiovanni, S. Olivares,
G. Brida, and M. Genovese, Experimental Realization of Quantum Illumination,
Phys. Rev. Lett. 110, 153603 (2013).

\bibitem{26} S. H. Tan, B. I. Erkmen, V. Giovannetti, S. Guha, S. Lloyd, L.
Maccone, S. Pirandola, and J. H. Shapiro, Quantum Illumination with Gaussian
States, Phys. Rev. Lett. 101, 253601 (2008).

\bibitem{27} Y. Guo, W. Ye, H. Zhong, and Q. Liao, Continuous-variable
quantum key distribution with non-Gaussian quantum catalysis, Phys. Rev. A
99, 032327 (2019).

\bibitem{28} W. Ye, H. Zhong, Q. Liao, D. Huang, L. Y. Hu, and Y. Guo,
Improvement of self-referenced continuous-variable quantum key distribution
with quantum photon catalysis, Opt. Express 27, 17186-17198 (2019).

\bibitem{29} W. Ye, Y. Guo, Y. Xia, H. Zhong, H. Zhang, J. Z. Ding, and L.Y.
Hu, Discrete modulation continuous-variable quantum key distribution based
on quantum catalysis. Acta Phys. Sin. 69, 060301 (2020).

\bibitem{30} Y. J. Zhao, Y. C. Zhang, B. J. Xu, S. Yu, and H. Guo,
Continuous-variable measurement-device-independent quantum key distribution
with virtual photon subtraction, Phys. Rev. A 97, 042328 (2018).

\bibitem{31} H. X. Ma, P. Huang, D. Y. Bai, S. Y. Wang, W. S. Bao, and G. H.
Zeng, Continuous-variable measurement-deviceindependent quantum key
distribution with photon subtraction, Phys. Rev. A 97, 042329 (2018).

\bibitem{32} Y. Yang and F. L. Li, Entanglement properties of non-Gaussian
resources generated via photon subtraction and addition and
continuous-variable quantum-teleportation improvement, Phys. Rev. A 80,
022315 (2009).

\bibitem{33} T. Opatrny, G. Kurizki, and D. G. Welsch, Improvement on
teleportation of continuous variables by photon subtraction via conditional
measurement, Phys. Rev. A 61, 032302 (2000).

\bibitem{34} S. Takeda, H. Benichi, T. Mizuta, N. Lee, J. Yoshikawa, and A.
Furusawa, Quantum mode filtering of non-Gaussian states for
teleportation-based quantum information processing, Phys. Rev. A 85, 053824
(2012).

\bibitem{35} A. Kitagawa, M. Takeoka, M. Sasaki, and A. Chefles,
Entanglement evaluation of non-Gaussian states generated by photon
subtraction from squeezed states, Phys. Rev. A 73, 042310 (2006).

\bibitem{36} Y. Yang and F. L. Li, Nonclassicality of photon-subtracted and
photon-added-then-subtracted Gaussian states, J. Opt. Soc. Am. B \textbf{26}
000830 (2009).

\bibitem{37} M. S. Kim, H. Jeong, A. Zavatta, V. Parigi, and M. Bellini,
Scheme for Proving the Bosonic Commutation Relation Using Single-Photon
Interference, Phys. Rev. Lett. 101, 260401 (2008).

\bibitem{38} H. Zhang, W. Ye, Y. Xia, S. K. Chang, C. P. Wei, and L. Y. Hu,
Improvement of the entanglement properties for entangled states using a
superposition of number-conserving operations,\ Laser Phys. Lett. 16, 085204
(2019).

\bibitem{39} S. D. Himadri, C. Arpita, and G. Rupamanjari, Generating
continuous variable entangled states for quantum teleportation using a
superposition of number-conserving operations, J. Phys. B: At. Mol. Opt.
Phys. 48, 185502 (2015).

\bibitem{40} S. Ataman, Optimal Mach-Zehnder phase sensitivity with Gaussian
states, Phys. Rev. A 100, 063821 (2019).

\bibitem{41} L. L. Guo, Y. F. Yu, Z. M. Zhang, Improving the phase
sensitivity of an SU(1,1) interferometer with photon-added squeezed vacuum
light, Opt. Express 26, 29099 (2018).

\bibitem{42} X. Y. Hu, C. P. Wei, Y. F. Yu, and Z. M. Zhang, Enhanced phase
sensitivity of an SU(1,1) interferometer with displaced squeezed vacuum
light, Front. Phys. 11, 114203 (2016).

\bibitem{43} D. Li, B. T. Gard, Y. Gao, C. H. Yuan, W. P. Zhang, H. Lee, and
J. P. Dowling, Phase sensitivity at the Heisenberg limit in an SU(1,1)
interferometer via parity detection, Phys. Rev. A 94, 063840 (2016).

\bibitem{44} R. A. Campos, C. C. Gerry, and A. Benmoussa, Optical
interferometry at the Heisenberg limit with twin Fock states and parity
measurements, Phys. Rev. A 68, 023810 (2003).

\bibitem{45} T. Kim, O. Pfister, M. Holland, J. Noh, and J. Hall, Influence
of decorrelation on Heisenberg-limited interferometry with quantum
correlated photons, Phys. Rev. A 57, 4004 (1998).

\bibitem{46} C. C. Gerry, Heisenberg-limit interferometry with four-wave
mixers operating in a nonlinear regime, Phys. Rev. A 61, 043811 (2000).

\bibitem{47} C. C. Gerry and R. A. Campos, Generation of maximally entangled
photonic states with a quantum-optical Fredkin gate, Phys. Rev. A 64, 063814
(2001).

\bibitem{48} A. Joshia and S.V. Lawande, Properties of Squeezed Binomial
States and Squeezed Negative Binomial States, J. Mod. Opt. 38, 2009 (1991).

\bibitem{49} C. T. Lee, Many-photon anti-bunching in generalized pair
coherent states, Phys. Rev. A 41, 1569 (1990).

\bibitem{50} W. Ye, K. Z. Zhang, H. L. Zhang, X. X. Xu, and L. Y. Hu,
Laguerre-polynomial-weighted squeezed vacuum: generation and its properties
of entanglement, Laser Phys. Lett. \textbf{15 }025204 (2018).

\bibitem{51} B. Yurke, S. L. McCall, and J. R. Klauder, SU(2) and SU(1,1)
interferometers, Phys. Rev. A 33, 4033 (1986).

\bibitem{52} Y. M. Zhang, X. W. Li, W. Yang, and G. R. Jin, Quantum Fisher
information of entangled coherent states in the presence of photon loss,
Phys. Rev. A 88, 043832 (2013).

\bibitem{53} R. Birrittella, J. Mimih, and C. C. Gerry, Multiphoton quantum
interference at a beam splitter and the approach to Heisenberg-limited
interferometry, Phys. Rev. A 86, 063828 (2012).

\bibitem{54} P. R. Bevinglon, Data Reduction and Error Analysis for the
Physical Sciences, McGraw-Hill, NewYork, 1969.

\bibitem{55} K. P. Seshadreesan, S. Kim, J. P. Dowling, and H. Lee, Phase
estimation at the quantum Cram-Rao bound via parity detection, Phys. Rev. A
87 043833 (2013).

\bibitem{56} T. W. Lee, S. D. Huver, H. Lee, L. Kaplan, S. B. McCracken, C.
Min, D. B. Uskov, C. F. Wildfeuer, G. Veronis, and J. P. Dowling,
Optimization of quantum interferometric metrological sensors in the presence
of photon loss, Phys. Rev. A 80, 063803 (2009).

\bibitem{57} D. Li, C. H. Yuan, Y. Yao, W. Jiang, M. Li, and W. P. Zhang,
Effects of loss on the phase sensitivity with parity detection in an SU(1,1)
interferometer, J. Opt. Soc. Am. B, 35 001080 (2018).

\bibitem{58a} M. O. Scully and M. S. Zubairy, Quantum Optics (Cambridge:
Cambridge University Press 1997).

\bibitem{58} H. Y. Fan and H. R. Zaidi, Application of IWOP technique to the
generalized Weyl correspondence, Phys. Lett. A 124, 303, (1987).

\bibitem{59} H. Y. Fan, Newton-Leibniz integration for ket-bra operators in
quantum mechanics (V)-Deriving normally ordered
bivariate-normal-distribution form of density operators and developing their
phase space formalism, Ann. Phys. \textbf{323}, 1502 (2008).

\bibitem{60} H. Y. Fan, H. L. Lu, Y. Fan, Newton-Leibniz integration for
ket-bra operators in quantum mechanics and derivation of entangled state
representations, Ann. Phys. \textbf{321}, 480 (2006).
\end{thebibliography}
\end{document}